\documentclass[aps,english,preprintnumbers,nofootinbib,twocolumn,physics]{revtex4-1}

\usepackage{booktabs}
\usepackage{amsfonts,amsmath,amssymb}
\usepackage{graphicx}
\usepackage[utf8]{inputenc}
\usepackage{hyperref}
\usepackage{babel}
\usepackage{natbib}
\usepackage{xcolor}
\usepackage{placeins}
\hypersetup{hidelinks}

\usepackage{mathalpha}
\usepackage{mathrsfs}
\usepackage{comment}

\begin{document}

\title{Krylov-Space Memory Cores}
\author{Mohsen Alishahiha
and  
Mohammad Javad Vasli}
\email{alishah@ipm.ir,  vasli@ipm.ir}
\affiliation{School of Quantum Physics and Matter, 
Institute for Research in Fundamental 
Sciences (IPM),\\
P.O. Box 19395-5531, Tehran, Iran
}

\begin{abstract}

We introduce Krylov-space memory cores as stationary, depth-resolved structures that reveal how anomalous initial-state memory is organized inside the Krylov 
space of otherwise thermalizing nonintegrable systems.
The stationary occupation profile identifies where late-time probability
is concentrated along the Krylov chain, while complementary diagnostics 
of residual equilibration fluctuations, deviation from the Gibbs reference, and long-time Krylov-current fluctuations determine the physical character 
of that region. 
Across weak thermalization, confinement-induced anomalous dynamics, and many-body scarring, anomalous initial states develop compact low-depth memory cores that carry appreciable residual fluctuations, Gibbs mismatch, and persistent current-fluctuation activity. These cores are often embedded within substantially broader stationary occupation halos.
The reference states considered, by contrast, do not exhibit a comparable combination of signal strength and spatial compactness. An auxiliary integrable comparison further shows that compact Krylov memory is state selective rather than a generic consequence of integrability. Krylov-space memory cores therefore provide a stationary framework for identifying where structured quantum memory resides and how it remains dynamically encoded.

\end{abstract}

\maketitle

\section{Introduction}

Understanding how isolated quantum many-body systems equilibrate and
thermalize remains a central problem in nonequilibrium quantum physics.
Although the dynamics of a closed system is unitary and reversible, many
interacting systems approach late-time states in which selected observables
are well described by statistical ensembles. The eigenstate thermalization
hypothesis (ETH) provides a microscopic explanation for this behavior in
generic many-body systems~\cite{Deutsch:1991,Srednicki:1994mfb}, and has led
to close connections among thermalization, quantum chaos, spectral
statistics, entanglement growth, and information
scrambling~\cite{Bohigas1984,rigol2008thermalization,DAlessio2016}; see
Refs.~\cite{Mori2018,Alishahiha:2025rdg} for reviews.

An important exception occurs when selected initial states display anomalous
relaxation even though the Hamiltonian is nonintegrable and generic states
thermalize. Weak thermalization~\cite{Banuls:2010zki}, confinement-induced
anomalous dynamics~\cite{McCoyWu1978,Kormos2017Confinement,
James2019Confinement}, and quantum many-body
scars~\cite{Bernien:2017ubn,turner2018quantum,turner2018weak,
shiraishi2017systematic,mori2017thermalization,Moudgalya:2021xlu} are
prominent examples. These phenomena are commonly identified through local
observables~\cite{rigol2008thermalization,DAlessio2016}, entanglement
measures~\cite{Calabrese2005,Eisert2010}, spectral
statistics~\cite{Bohigas1984,Haake2010}, and complexity-based
probes~\cite{Alishahiha:2024rwm,Menzler:2024ifs}. Such diagnostics establish
that the dynamics is atypical, but they do not directly reveal whether the
surviving late-time information acquires a common spatial organization inside
the state-dependent dynamical subspace.

Krylov space~\cite{Lanczos1950,
Haydock1980,ParkLight1986,
viswanath1994recursion} provides a natural setting for this question. Given a
Hamiltonian $H$ and an initial state $|\psi_0\rangle$, the Lanczos algorithm
generates an ordered orthonormal basis $\{|n\rangle\}$ spanning the cyclic
subspace
\begin{equation}
\mathcal K
=
{\rm span}
\left\{
|\psi_0\rangle,
H|\psi_0\rangle,
H^2|\psi_0\rangle,
\ldots
\right\}.
\end{equation}
In this basis, the Schr\"odinger equation becomes an effective
one-dimensional hopping problem, and the Krylov index defines a
state-dependent depth along the resulting chain~\cite{Parker:2018yvk,
Rabinovici:2020ryf,Balasubramanian:2022tpr,
PhysRevResearch.4.013041,Barbon:2019wsy,Alishahiha:2022anw}. This emergent
geometry has become a useful tool for studying operator growth, quantum
chaos, information scrambling, and complexity growth~\cite{Dymarsky:2021bjq,Bhattacharjee:2022vlt,Avdoshkin:2022xuw,
Camargo:2022rnt,Vasli:2023syq,Imani:2025etp,Rabinovici:2022beu,
Scialchi:2023bmw,Trigueros:2021rwj,Espanol:2022cqr,Erdmenger:2023wjg,
Huh:2023jxt,Camargo:2024deu,
Nandy:2024wwv,Bhattacharjee:2024yxj,
Balasubramanian:2024ghv,Baggioli:2024wbz,
Alishahiha:2024vbf,
FarajiAstaneh:2025rlc,Bhattacharya:2024szw,
Huh:2024ytz,Baggioli:2025ohh,
FarajiAstaneh:2025thi,
Roychowdhury:2026vzq,Banerjee:2026kav}; see
Refs.~\cite{Nandy:2024evd,Rabinovici:2025otw} for reviews.
Beyond growth-based measures, fine structure in the Lanczos coefficients can itself encode low-energy spatial organization; for boundary operators in long-range Kitaev chains, a Krylov-staggering diagnostic was recently shown to track whether the lowest excitation gap is governed by edge-localized or bulk-extended modes~\cite{JhaMenzler2026KitaevKrylov}.

Much of the literature has focused on Krylov complexity~\cite{Parker:2018yvk},
\begin{equation}
{\cal C}(t)
=
\sum_{n=0}^{\mathcal D_0-1} n P_n(t),
\qquad
P_n(t)=|\varphi_n(t)|^2,
\end{equation}
where $|\psi(t)\rangle=\sum_n\varphi_n(t)|n\rangle$ and $\mathcal D_0$ is the
Krylov-space dimension. Krylov complexity records only the mean depth of the
wavepacket. The full depth-resolved profile contains additional information
and can reveal localization, dynamical constraints, and nonergodic structure
that are not captured by its first moment~\cite{Trigueros:2021rwj,
Bhattacharjee:2021jce,Nandy:2024evd,Menzler:2024ifs,
Pain:2026krylovAnatomy}. Recent work has also begun to study Krylov
distributions beyond Krylov complexity~\cite{Alishahiha:2026fnu,
Alishahiha:2026qzc}.

The present work builds on these developments and on our earlier investigation
of thermalization in the Krylov basis~\cite{Alishahiha:2024rwm}. We develop a
stationary, depth-resolved framework for anomalous initial-state dynamics. The
stationary occupation profile $\Pi_n$ is the diagonal ensemble expressed in
the Krylov basis and describes the late-time probability background along the
chain. On this background we study three complementary quantities: the
residual fluctuation profile $\overline{\chi_n}$, the Gibbs-deviation profile
$d_n$, and the bond-resolved Krylov-current fluctuation activity $\Gamma_n$.
They quantify, respectively, long-time fluctuations of Krylov occupations,
the deviation of the stationary profile from the Gibbs reference, and the
variance of the probability current through neighboring Krylov bonds. The
last quantity is a central new ingredient of the construction because it
resolves whether the stationary dynamics continues to support probability
exchange after the mean current has vanished.

The principal result of this work is that these three diagnostics reveal a
recurring stationary structure for atypical initial states in otherwise
thermalizing systems. At the accessible system sizes, weakly thermalizing,
confinement-sensitive, and scarred initial states exhibit appreciable
fluctuation, Gibbs-deviation, and current-activity weights concentrated within
a common compact region near the beginning of the Krylov chain. Typical
thermalizing reference states evolving under the same Hamiltonians do not
show a comparable combination: their residual weights are weak, broadly
distributed, or one of the three components is absent. We call the compact,
dynamically active region selected by the anomalous states a
\emph{Krylov-space memory core}\footnote{Throughout this work, the term ``memory'' is used operationally to denote
state-selective stationary residual structure that remains fluctuating,
mismatched with the equilibrium benchmark, and dynamically active in the
Krylov basis. It is not intended as a measure of information-storage
capacity, and the three diagnostics are not claimed to be unique or
exhaustive.}. The repeated appearance of this structure in
three models with different microscopic mechanisms is the main achievement of
the paper.

The memory core is defined by the spatial concentration of the 
residual fluctuation, Gibbs-mismatch, and current-activity weights, not by the stationary occupation 
profile alone. The occupation profile supplies the probability 
background on which
the other quantities are supported, but it does not determine whether 
that
background remains fluctuating, differs from the equilibrium reference, 
or
supports long-time current fluctuations. We therefore characterize a memory
core through the total residual weights and the cumulative depth scales of
$\overline{\chi_n}$, $d_n^2$, and $\Gamma_n$. Their common low-depth
concentration is established by the numerical results rather than imposed by
the formal definitions.

Our claim is specifically about state-selective anomalous dynamics in
nonintegrable systems that otherwise thermalize. The memory core is not
proposed as a criterion for distinguishing integrable from nonintegrable
Hamiltonians. An auxiliary integrable Ising comparison shows that structured
initial states can also exhibit compact residual profiles, while a
symmetry-matched random product state need not. Thus the occurrence of a core
depends on the pair $(H,|\psi_0\rangle)$, while its significance in the main
examples lies in the sharp contrast between anomalous and typical
thermalizing states under the same nonintegrable Hamiltonian.

We also compare the finite many-body systems with exactly solvable infinite
Krylov chains. In these escaping geometries, probability continues to move
toward arbitrarily large depth and no normalizable stationary fixed-depth
profile remains. This comparison separates stationary memory from Krylov
growth: rapid spreading alone does not generate a memory core.

The remainder of the paper is organized as follows. In
Sec.~\ref{sec:stationary} we introduce the stationary occupation profile and
the three complementary Krylov diagnostics, together with their cumulative
characterization. Section~\ref{sec:numerics} presents the numerical results
for the chaotic Ising chain, the confinement regime, and the PXP model. In
Sec.~\ref{sec:escaping_geometries} we compare these results with exactly
solvable escaping Krylov geometries. Section~\ref{sec:conclusions} summarizes
our conclusions. The integrable comparison, finite-size results, and
technical details are collected in the Appendices.



\section{Stationary Krylov diagnostics}
\label{sec:stationary}

We first establish the Krylov-space framework used throughout the paper.
Consider a finite-dimensional quantum system with a time-independent
Hamiltonian $H$ acting on a Hilbert space $\mathcal H$ of dimension
$\mathcal D$. Given a normalized initial state $|\psi_0\rangle$, its cyclic
Krylov space is
\begin{equation}
\mathcal K
=
{\rm span}\!
\left\{
|\psi_0\rangle,
H|\psi_0\rangle,
H^2|\psi_0\rangle,
\ldots
\right\},
\label{eq:Krylov-space}
\end{equation}
with dimension $\mathcal D_0\leq\mathcal D$. The subspace $\mathcal K$ is
invariant under $H$ and contains the complete unitary orbit generated from
$|\psi_0\rangle$.

An ordered orthonormal basis
$\{|n\rangle\}_{n=0}^{\mathcal D_0-1}$ of $\mathcal K$ is generated by the
Lanczos algorithm~\cite{Lanczos1950,viswanath1994recursion}. Starting from
$|0\rangle=|\psi_0\rangle$ and $|-1\rangle=0$, one defines
\begin{align}
|\widehat{n+1}\rangle
&=
(H-a_n)|n\rangle-b_n|n-1\rangle,
\\
|n+1\rangle
&=
\frac{|\widehat{n+1}\rangle}{b_{n+1}},
\label{eq:Lanczos-recursion}
\end{align}
where
\begin{equation}
a_n=\langle n|H|n\rangle,
\qquad
b_{n+1}=\sqrt{\langle\widehat{n+1}|\widehat{n+1}\rangle}.
\end{equation}
The restriction $H_K\equiv H|_{\mathcal K}$ then has the tridiagonal form
\begin{equation}
H_K
=
\sum_{n=0}^{\mathcal D_0-1}a_n|n\rangle\langle n|
+
\sum_{n=0}^{\mathcal D_0-2}b_{n+1}
\left(
|n\rangle\langle n+1|+|n+1\rangle\langle n|
\right).
\label{eq:tridiagonal-H}
\end{equation}

The evolved state is expanded as
\begin{equation}
|\psi(t)\rangle
=
\sum_{n=0}^{\mathcal D_0-1}\varphi_n(t)|n\rangle,
\qquad
\sum_{n=0}^{\mathcal D_0-1}|\varphi_n(t)|^2=1,
\end{equation}
and its amplitudes obey
\begin{equation}
i\partial_t\varphi_n(t)
=
b_{n+1}\varphi_{n+1}(t)
+a_n\varphi_n(t)
+b_n\varphi_{n-1}(t),
\label{eq:Sch}
\end{equation}
with $b_0=0$. The many-body evolution is therefore represented as hopping
on an emergent one-dimensional chain. The index $n$ defines a
state-dependent Krylov depth: small $n$ labels directions reached after only
a few actions of $H$ on the initial state, whereas larger $n$ probes
progressively deeper parts of the cyclic subspace.

The basic time-dependent quantity is the Krylov probability distribution
\begin{equation}
P_n(t)=|\varphi_n(t)|^2,
\label{eq:Pn}
\end{equation}
or, equivalently,
\begin{equation}
P_n(t)
=
\langle n|\rho(t)|n\rangle
=
{\rm Tr}\!\left[\rho(t)O_n\right],
\qquad
O_n=|n\rangle\langle n|,
\end{equation}
where $\rho(t)=|\psi(t)\rangle\langle\psi(t)|$. The projectors $O_n$ are
state dependent and generally nonlocal. The quantities introduced below are
therefore fine-grained Krylov-resolved diagnostics; they complement, rather
than replace, conventional few-body tests of equilibration and
thermalization.

Our aim is to distinguish stationary probability from the physical content
carried by that probability. A site may retain nonzero late-time occupation
while being nearly equilibrated, close to the chosen equilibrium reference,
and weakly involved in probability exchange. 
Moreover, the profiles may be
broad, tail dominated, multi-peaked, or concentrated near the Krylov origin.

\subsection{Active spectral measure and stationary occupation}

The stationary quantities are most cleanly formulated in terms of the
spectral measure selected by the initial state. To treat exact energy
degeneracies without choosing an arbitrary basis inside a degenerate
subspace, write
\begin{equation}
H
=
\sum_{E\in {\rm spec}(H)} E\,\mathsf P_E,
\end{equation}
where $\mathsf P_E$ projects onto the complete eigenspace of energy $E$. The
set of distinct energies active in the evolution is
\begin{equation}
\mathcal E_0
=
\left\{
E\in {\rm spec}(H):
\mathsf P_E|\psi_0\rangle\neq0
\right\}.
\label{eq:active-spectrum}
\end{equation}
For each $E\in\mathcal E_0$, define
\begin{equation}
\omega_E
=
\langle\psi_0|\mathsf P_E|\psi_0\rangle,
\qquad
|\Phi_E\rangle
=
\frac{\mathsf P_E|\psi_0\rangle}{\sqrt{\omega_E}},
\qquad
\omega_E>0.
\label{eq:active-spectral-data}
\end{equation}
The vectors $|\Phi_E\rangle$ are orthonormal for distinct active energies.
Repeated action by $H$ only multiplies the component $|\Phi_E\rangle$ by
powers of $E$; conversely, polynomial interpolation on the finite set
$\mathcal E_0$ isolates each active component. Hence
\begin{equation}
\mathcal K
=
{\rm span}\left\{|\Phi_E\rangle:E\in\mathcal E_0\right\},
\qquad
\mathcal D_0=|\mathcal E_0|,
\label{eq:K-active-spectrum}
\end{equation}
and
\begin{equation}
H_K
=
\sum_{E\in\mathcal E_0}E|\Phi_E\rangle\langle\Phi_E|.
\label{eq:HK-active-spectrum}
\end{equation}
Thus an exactly degenerate energy eigenspace contributes one direction to the
cyclic space: the direction selected by the initial state. Zero-energy
subspaces are included in precisely the same way.

In this representation,
\begin{equation}
|\psi(t)\rangle
=
\sum_{E\in\mathcal E_0}
\sqrt{\omega_E}\,e^{-iEt}|\Phi_E\rangle,
\end{equation}
and dephasing between distinct energies gives
\begin{equation}
\rho_{\rm diag}
=
\sum_{E\in\mathcal E_0}
\omega_E|\Phi_E\rangle\langle\Phi_E|.
\label{eq:rho-diag}
\end{equation}
Equation~\eqref{eq:rho-diag} remains exact in the presence of energy-level
degeneracies: coherence inside each degenerate eigenspace is retained through
the single active vector $|\Phi_E\rangle$.

For finite $\mathcal D_0$, the infinite-time averaged occupation at Krylov
depth $n$ is
\begin{align}
\Pi_n
&=
\overline{P_n(t)}
=
\lim_{T\to\infty}\frac{1}{T}\int_0^T P_n(t)\,dt
\nonumber\\
&=
\langle n|\rho_{\rm diag}|n\rangle
=
\sum_{E\in\mathcal E_0}
\omega_E|\langle n|\Phi_E\rangle|^2.
\label{eq:Pi-def}
\end{align}
It is a normalized stationary probability profile,
\begin{equation}
\sum_{n=0}^{\mathcal D_0-1}\Pi_n=1,
\label{eq:Pi-normalization}
\end{equation}
and should not be confused with a thermal distribution. Since
$|0\rangle=|\psi_0\rangle$,
\begin{equation}
\Pi_0
=
\sum_{E\in\mathcal E_0}\omega_E^2,
\end{equation}
which is the inverse participation ratio over the distinct active energy
eigenspaces.

The full sequence $\{\Pi_n\}$ describes the stationary probability
background along the Krylov chain. It tells us where probability remains
after dephasing, but not whether that probability carries a sizable residual stationary structure. The latter requires information about temporal fluctuations,
equilibrium mismatch, and phase-sensitive exchange, which we now introduce.

\subsection{Three complementary stationary diagnostics}

Having established the stationary occupation profile, we now introduce three
complementary depth-resolved quantities that characterize the residual
stationary structure and dynamical activity along the Krylov chain. These
quantities are defined on the same state-dependent Krylov basis and will
subsequently be analyzed through their local profiles, total weights, and
cumulative distributions.

For the compact spectral expressions used below, we impose the
nondegenerate-active-gap condition in which for $E,E',F,F'\in\mathcal E_0$
and $E\neq E', F\neq F'$, one has
\begin{equation}
\begin{aligned}
E-E'&=F-F',
\quad
\Longrightarrow\quad
E=F,
\;\;\;
E'=F'.
\end{aligned}
\label{eq:nondegenerate-active-gaps}
\end{equation}
Thus every nonzero energy difference between distinct active energies is
unique. This assumption concerns degeneracies of energy \emph{gaps}, not
degeneracies of the energy eigenspaces themselves. Exact level degeneracies,
including possible zero-energy eigenspaces, have already been incorporated
through the active spectral projectors, weights, and vectors introduced in
Eqs.~\eqref{eq:active-spectrum}--\eqref{eq:active-spectral-data}.

When Eq.~\eqref{eq:nondegenerate-active-gaps} is satisfied, the relevant
infinite-time averages reduce to the compact spectral expressions given
below. If the condition is violated, the definitions of the diagnostics
remain unchanged, but the long-time averages must retain all contributions
associated with equal gaps, or equivalently be evaluated directly from the
time-dependent quantities.

\subsubsection{Residual equilibration fluctuations}

A stationary mean does not imply that the corresponding observable has
settled close to that mean. To resolve the remaining temporal motion of the
Krylov occupation, define
\begin{equation}
\chi_n(t)
=
\left(P_n(t)-\Pi_n\right)^2,
\end{equation}
and
\begin{equation}
\overline{\chi_n}
=
\overline{\left(P_n(t)-\Pi_n\right)^2}.
\label{eq:chi-def}
\end{equation}
Thus $\overline{\chi_n}$ is the long-time variance of the Krylov projector
$O_n$. A small value means that the occupation at depth $n$ remains close to
its stationary value, whereas a large value signals persistent
Krylov-resolved fluctuations.

Under Eq.~\eqref{eq:nondegenerate-active-gaps},
\begin{equation}
\overline{\chi_n}
=
\sum_{\substack{E,E'\in\mathcal E_0\\E\neq E'}}
\omega_E\omega_{E'}
|\langle n|\Phi_E\rangle|^2
|\langle n|\Phi_{E'}\rangle|^2.
\label{eq:chi-spectral}
\end{equation}
Equivalently,
\begin{equation}
\overline{\chi_n}
=
\Pi_n^2
-
\sum_{E\in\mathcal E_0}
\omega_E^2|\langle n|\Phi_E\rangle|^4,
\label{eq:chi-Pi-relation}
\end{equation}
so that
\begin{equation}
0\leq\overline{\chi_n}\leq\Pi_n^2.
\label{eq:chi-local-bound}
\end{equation}
The occupation therefore supplies the probability needed for a fluctuation,
but it does not determine whether that probability fluctuates. For fixed
$\Pi_n$, the subtraction term in Eq.~\eqref{eq:chi-Pi-relation} depends on
how the local stationary weight is distributed among the active energies. A
single dominant energy component suppresses the variance, whereas a more
mixed local spectral composition can produce a larger fluctuation signal.

A useful state-wide bound is
\begin{equation}
\overline{\chi_n}
\leq
\Pi_n^2
\leq
{\rm Tr}(\rho_{\rm diag}^2)
=
\sum_{E\in\mathcal E_0}\omega_E^2
=
\Pi_0.
\label{eq:chi-bound}
\end{equation}
These bounds constrain the available scale but neither guarantee an
appreciable total fluctuation weight nor determine where that weight is
concentrated along the chain.

\subsubsection{Krylov-resolved Gibbs deviation}

Equilibration and thermalization answer different questions. The former
concerns temporal fluctuations around a stationary state; the latter concerns
whether that stationary state agrees with an appropriate equilibrium
reference. For the nonintegrable systems studied in the main text, we use the
Gibbs state of the cyclic Hamiltonian,
\begin{equation}
\rho_{\rm Gibbs}
=
\frac{e^{-\beta H_K}}{Z_K(\beta)},
\qquad
Z_K(\beta)
=
{\rm Tr}_{\mathcal K}\!\left(e^{-\beta H_K}\right)
=
\sum_{E\in\mathcal E_0}e^{-\beta E},
\label{eq:gibbs}
\end{equation}
where $\beta$ is fixed by energy matching,
\begin{equation}
{\rm Tr}_{\mathcal K}\!\left(\rho_{\rm Gibbs}H_K\right)
=
\langle\psi_0|H|\psi_0\rangle.
\label{eq:beta-match}
\end{equation}
Equivalently,
\begin{equation}
\rho_{\rm Gibbs}
=
\sum_{E\in\mathcal E_0}
g_E|\Phi_E\rangle\langle\Phi_E|,
\qquad
g_E=\frac{e^{-\beta E}}{Z_K(\beta)}.
\label{eq:gibbs-active-basis}
\end{equation}
When $\mathcal K$ coincides with the relevant physical symmetry or
constrained sector, this is the usual Gibbs ensemble in that sector. When
$\mathcal K$ is a proper cyclic subspace, it is a state-dependent
cyclic-space equilibrium benchmark.

The corresponding Krylov occupation is
\begin{equation}
\Pi_n^{({\rm Gibbs})}
=
\langle n|\rho_{\rm Gibbs}|n\rangle
=
\sum_{E\in\mathcal E_0}
g_E|\langle n|\Phi_E\rangle|^2,
\label{eq:gibbs-profile}
\end{equation}
and we define the signed deviation
\begin{equation}
d_n
=
\Pi_n-\Pi_n^{({\rm Gibbs})}
=
\langle n|\left(\rho_{\rm diag}-\rho_{\rm Gibbs}\right)|n\rangle.
\label{eq:d-profile}
\end{equation}
The role of $d_n$ is not to locate stationary probability independently of
$\Pi_n$, but to identify where that stationary probability carries an
equilibrium-reference mismatch. An occupied region may be reproduced almost
exactly by the Gibbs profile and hence have $d_n\simeq0$, or it may retain a
large signed excess or deficit. In the cumulative analysis below, $d_n^2$ is
used as the associated nonnegative mismatch weight.

Because the Krylov projectors are generally nonlocal and state dependent,
$d_n$ is a fine-grained ensemble-comparison diagnostic rather than a
conventional local-observable test of thermalization. In an integrable model,
the ordinary Gibbs state is generally not the appropriate stationary
reference; unless it is replaced by a generalized Gibbs ensemble, $d_n$
should then be described only as an ordinary-Gibbs mismatch.

\subsubsection{Krylov-current fluctuation activity}

The previous two diagnostics are site resolved. They do not directly measure
whether probability continues to move between neighboring Krylov depths.
The nearest-neighbor form of Eq.~\eqref{eq:Sch} provides an exact continuity
equation,
\begin{equation}
\partial_tP_n(t)
=
J_n(t)-J_{n+1}(t),
\label{eq:continuity}
\end{equation}
with $J_0=J_{\mathcal D_0}=0$. The current through the bond joining sites
$n-1$ and $n$ is
\begin{equation}
J_n(t)
=
2b_n\,{\rm Im}\!\left[\varphi_n^*(t)\varphi_{n-1}(t)\right].
\label{eq:current}
\end{equation}
Unlike an occupation, $J_n(t)$ depends on the relative phase of neighboring
amplitudes. It therefore probes coherent probability exchange on the
emergent Krylov chain, not physical-space transport.\footnote{Related
geometric flow pictures appear in higher-dimensional Krylov-graph
constructions~\cite{Murugan:2026superadditivity}; here we consider the
one-dimensional chain generated from a single initial state.}

The same current controls the instantaneous growth of Krylov complexity,
\begin{equation}
\mathcal C(t)
=
\sum_{n=0}^{\mathcal D_0-1}nP_n(t),
\qquad
\frac{d\mathcal C}{dt}
=
\sum_{n=1}^{\mathcal D_0-1}J_n(t).
\label{eq:Cdot}
\end{equation}
In a finite isolated system the currents can repeatedly reverse direction.
Their infinite-time means vanish whenever the relevant averages exist,
without requiring Eq.~\eqref{eq:nondegenerate-active-gaps}. Indeed,
$P_n(t)$ is bounded, so $\overline{\partial_tP_n}=0$; averaging
Eq.~\eqref{eq:continuity} gives
$\overline{J_n}=\overline{J_{n+1}}$, and the boundary conditions imply
\begin{equation}
\overline{J_n(t)}=0
\end{equation}
for every bond.

Zero mean current does not imply that a bond is dynamically quiet. We define
its stationary current-fluctuation activity by
\begin{equation}
\Gamma_n
=
\overline{J_n^2(t)}.
\label{eq:Gamma-def}
\end{equation}
A large $\Gamma_n$ indicates persistent back-and-forth probability exchange.
Because the current is squared, $\Gamma_n$ contains no information about a
net direction and should not be interpreted as directed transport,
transport efficiency, or irreversible spreading.

For systems satisfying the nondegenerate-active-gap condition, and in the standard Lanczos
gauge in which $a_n$ and $b_n$ are real, the exact
gap-resolved expression reduces to 
(see Appendix~
\ref{app:gamma_derivation})
\begin{equation}
\Gamma_n
=
2b_n^2
\left[
\Pi_n\Pi_{n-1}
-
|\langle n|\rho_{\rm diag}|n-1\rangle|^2
\right].
\label{eq:GammaExact}
\end{equation}
For gap-degenerate spectra such as that of the PXP model, we instead evaluate the coherent gap-resolved expression in
Eq.~\eqref{eq:Gamma-gap-resolved-app}.

The neighboring occupations and the hopping $b_n$ set the probability and
coupling available on the bond, while the diagonal-ensemble coherence
controls how much of that available scale appears as current fluctuations.
Cauchy--Schwarz gives
\begin{equation}
\left|\langle n|\rho_{\rm diag}|n-1\rangle\right|^2
\leq
\Pi_n\Pi_{n-1},
\end{equation}
and hence
\begin{equation}
0\leq\Gamma_n\leq2b_n^2\Pi_n\Pi_{n-1}.
\label{eq:Gamma-bound}
\end{equation}
Thus occupation on both sides of the bond is necessary for activity, but is
not sufficient to determine it. If active gaps are degenerate,
Eq.~\eqref{eq:GammaExact} is replaced by the gap-resolved average or by
direct long-time evaluation of Eq.~\eqref{eq:Gamma-def}.

\subsection{Cumulative diagnostics and memory-core depth}
\label{sec:cumulative}

Local profiles can contain oscillations, long tails, or several features,
which makes their characteristic extent difficult to read from individual
points alone. We therefore use cumulative quantities to separate two pieces
of information: the total strength of each signal and the depth over which
that signal is accumulated. 

For reference, the complete stationary probability background is summarized
by
\begin{equation}
\eta_\Pi(n_c)
=
\sum_{n=0}^{n_c}\Pi_n,
\label{eq:eta-Pi}
\end{equation}
and
\begin{equation}
n_\Pi^{(q)}
=
\min\left\{n_c:\eta_\Pi(n_c)\geq q\right\},
\qquad 0<q<1.
\label{eq:n-Pi-q}
\end{equation}
The quantity $n_\Pi^{(q)}$ measures the depth required to contain a fraction
$q$ of the full stationary probability cloud. It is useful as a geometric
baseline, but it does not say whether the occupied sites fluctuate, remain
out of equilibrium, or exchange probability dynamically.

For the three memory-related diagnostics, one may define the total nonnegative
weights
\begin{equation}
\mathcal X
=
\sum_{n=0}^{\mathcal D_0-1}\overline{\chi_n},
\qquad
\mathcal N
=
\sum_{n=0}^{\mathcal D_0-1}d_n^2,
\qquad
\mathcal G
=
\sum_{n=1}^{\mathcal D_0-1}\Gamma_n,
\label{eq:total-weights}
\end{equation}
and their partial sums
\begin{equation}
\mathcal X(n_c)
=
\sum_{n=0}^{n_c}\overline{\chi_n},
\qquad
\mathcal N(n_c)
=
\sum_{n=0}^{n_c}d_n^2,
\qquad
\mathcal G(n_c)
=
\sum_{n=1}^{n_c}\Gamma_n.
\label{eq:Wtruncated}
\end{equation}
For nonzero total weight, the normalized cumulative profiles are
\begin{equation}
\eta_\chi(n_c)
=
\frac{\mathcal X(n_c)}{\mathcal X},
\quad
\eta_d(n_c)
=
\frac{\mathcal N(n_c)}{\mathcal N},
\quad
\eta_\Gamma(n_c)
=
\frac{\mathcal G(n_c)}{\mathcal G}.
\label{eq:eta}
\end{equation}
Their threshold depths are
\begin{equation}
n_\ell^{(q)}
=
\min\left\{n_c:\eta_\ell(n_c)\geq q\right\},
\qquad
\ell=\chi,d,\Gamma.
\label{eq:nq}
\end{equation}
For $\ell=\Gamma$, $n_c$ labels a bond depth; comparisons with the two site
profiles are understood up to the natural one-step offset.

A small $n_\ell^{(q)}$ means that a fraction $q$ of the corresponding weight
is concentrated near the beginning of the Krylov chain. This depth must
always be interpreted together with the total weight $\mathcal X$,
$\mathcal N$, or $\mathcal G$. A rapidly saturating profile with a
numerically negligible total weight describes only a weak localized signal,
not a pronounced memory feature.

The term \emph{memory core} is intended to identify the region where
stationary probability acquires memory-bearing physical content, rather than
the full region in which stationary probability is present. We therefore
define, for a fixed threshold $q$,
\begin{equation}
n_{\rm mc}^{(q)}
=
\max\left\{
 n_\chi^{(q)},
 n_d^{(q)},
 n_\Gamma^{(q)}
\right\},
\label{eq:nmc}
\end{equation}
provided that $\mathcal X$, $\mathcal N$, and $\mathcal G$ are appreciable
relative to suitable reference states and are numerically resolved. This is
the smallest Krylov interval containing fraction q of each active residual weight. 

In the main-text comparisons we use $q=0.95$ and write
$n_{\rm mc}\equiv n_{\rm mc}^{(0.95)}$ when no ambiguity arises.  This
choice provides a compact primary summary of the region containing the
dominant part of each residual weight.  To test sensitivity to the
remaining tails, Appendix~\ref{app:core_scaling} repeats the finite-size
analysis at the stricter threshold $q=0.99$ and reports
$n_{\Pi}^{(0.99)}$, $n_{\chi}^{(0.99)}$, $n_d^{(0.99)}$, and
$n_{\Gamma}^{(0.99)}$, together with
\begin{equation}
 n_{\rm mc}^{(0.99)}
 =
 \max\left\{
 n_{\chi}^{(0.99)},
 n_d^{(0.99)},
 n_{\Gamma}^{(0.99)}
 \right\}.
 \label{eq:nmc_099_definition}
\end{equation}
The $99\%$ depths are used as a tail-sensitivity check rather than as a
second, competing definition of the memory core.  A robust core--halo
separation should remain visible when the threshold is increased from
$0.95$ to $0.99$, even though every cumulative depth is necessarily
nondecreasing with $q$.

The exclusion of $n_\Pi^{(q)}$ from Eq.~\eqref{eq:nmc} is physical rather
than data driven. The occupation profile supplies the stationary background
required for all three diagnostics, but the word ``memory'' refers to the
additional content resolved by $\overline{\chi_n}$, $d_n^2$, and $\Gamma_n$:
residual temporal fluctuations, equilibrium-reference mismatch, and
persistent current-fluctuation activity. Whether $n_\Pi^{(q)}$ is comparable
to, or larger than $n_{\rm mc}^{(q)}$ is a separate empirical
question. If the two scales are comparable, the memory-bearing structure
occupies most of the stationary probability cloud. If the occupation extends
substantially farther, the additional occupied region may be described as a
stationary halo that is comparatively weak with respect to the three
diagnostics used here.

Equation~\eqref{eq:nmc} defines a common enclosing interval; it does not imply
that the three local profiles are pointwise identical. Their detailed overlap
must be assessed from the local profiles or from a separate overlap analysis.
Likewise, the occupation-dependent bounds show that the diagnostics are
complementary rather than mathematically independent. What must be
established dynamically is that all three total weights are appreciable and
that their cumulative profiles select a comparable shallow interval.

The formalism therefore provides an operational test, not a prediction that a
core must exist. We use the term \emph{Krylov-space memory core} only when the
data exhibit appreciable residual fluctuation, mismatch from the equilibrium
reference, and current-fluctuation weights concentrated within a common low-depth region.
A normalizable stationary occupation profile is required for this stationary
construction, but neither a compact occupation profile nor any fixed relation
between $n_\Pi^{(q)}$ and $n_{\rm mc}^{(q)}$ is assumed.





\section{Krylov-space memory geometry: numerical evidence}
\label{sec:numerics}

We now apply the stationary framework of Sec.~\ref{sec:stationary} to
three nonintegrable many-body systems. The analysis does not assume that
a memory core exists. For each initial state, the stationary occupation
$\Pi_n$ and its cumulative depth $n_\Pi^{(q)}$ characterize the full
late-time probability background. The three residual profiles
$\overline{\chi_n}$, $d_n^2$, and $\Gamma_n$ are then used to determine
whether that background contains an appreciable memory-bearing region.

Throughout the main numerical section we use the primary threshold
$q=0.95$ and define
\begin{equation}
 n_{\rm mc}
 \equiv
 n_{\rm mc}^{(0.95)}
 =
 \max\left\{
 n_{\chi}^{(0.95)},
 n_d^{(0.95)},
 n_{\Gamma}^{(0.95)}
 \right\}.
 \label{eq:nmc_main_threshold}
\end{equation}
A pronounced memory core requires both appreciable total weights
$\mathcal X$, $\mathcal N$, and $\mathcal G$ relative to suitable
reference states and a small value of $n_{\rm mc}$ compared with the cyclic
dimension $\mathcal D_0$.  The occupation depth
$n_{\Pi}^{(0.95)}$ is reported separately.  Appendix~\ref{app:core_scaling}
reports the corresponding $99\%$ depths for the finite-size sequences and
thereby tests whether the inferred core--halo separation is stable against
the far tails of the four cumulative profiles.

We examine weak thermalization in a chaotic mixed-field Ising chain,
confinement-sensitive dynamics in a different regime of the same model,
and many-body scarring in the periodic PXP chain. The Gibbs-deviation
profile is used here as a fine-grained comparison between the diagonal
ensemble and the energy-matched Gibbs reference on the cyclic space. Since
the Krylov projectors are state dependent and generally nonlocal, this
comparison complements rather than replaces conventional few-body tests of
thermalization. Appendix~\ref{app:integrable_ising} provides an auxiliary
integrable comparison, where the ordinary Gibbs ensemble is not the
appropriate equilibrium reference.

\subsection{Chaotic mixed-field Ising chain}
\label{subsec:ising_num}

We first consider
\begin{equation}
H
=
-J\sum_{i=1}^{L-1}\sigma_i^z\sigma_{i+1}^z
-g\sum_{i=1}^{L}\sigma_i^x
-h\sum_{i=1}^{L}\sigma_i^z ,
\label{eq:ising_hamiltonian_num}
\end{equation}
with
\begin{equation}
J=1,
\qquad
g=-1.05,
\qquad
h=0.5,
\qquad
L=14.
\end{equation}
At these nonintegrable parameters, simple product states display markedly
different relaxation behavior under the same Hamiltonian
\cite{Banuls:2010zki}. We use the homogeneous states
\begin{equation}
|\theta,\varphi\rangle
=
\prod_{i=1}^{L}
\left(
\cos\frac{\theta}{2}|Z+\rangle_i
+
e^{i\varphi}\sin\frac{\theta}{2}|Z-\rangle_i
\right),
\label{eq:product_states_num}
\end{equation}
and, in particular,
\begin{equation}
|X+\rangle
=
\left|\frac{\pi}{2},0\right\rangle,
\qquad
|Y+\rangle
=
\left|\frac{\pi}{2},\frac{\pi}{2}\right\rangle,
\qquad
|Z+\rangle
=
|0,0\rangle .
\label{eq:xyz_states_num}
\end{equation}
The state $|Y+\rangle$ thermalizes efficiently, $|X+\rangle$ exhibits
weak thermalization, and $|Z+\rangle$ has intermediate behavior
\cite{Banuls:2010zki}.
We also include ten symmetry-matched random
product controls in the reflection-even sector. These controls provide a
benchmark for typical thermalizing dynamics without changing the symmetry
setting.

\begin{figure}[t]
\centering
\includegraphics[width=0.35\textwidth]{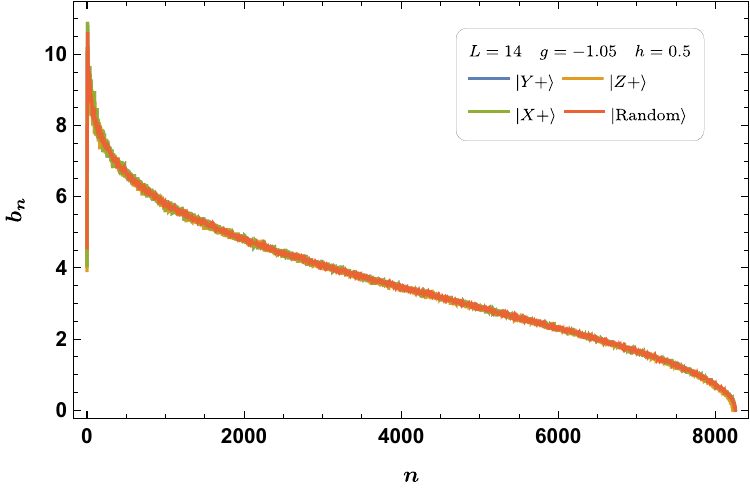}
\caption{
Lanczos coefficients $b_n$ for the chaotic mixed-field Ising chain
($L=14$, $J=1$, $g=-1.05$, $h=0.5$). All states are analyzed in the
reflection-even setting of dimension $8256$; their exact cyclic dimensions
are state dependent and are given in Table~\ref{tab:ising_total_weights}.
The similar large-depth envelopes show that the state-selective memory
structure is not determined by the overall growth of $b_n$ alone.
}
\label{fig:isingbn}
\end{figure}

The reflection-even sector has dimension $8256$. The exact cyclic
dimensions are state dependent and are listed in
Table~\ref{tab:ising_total_weights}. Figure~\ref{fig:isingbn} shows the
corresponding Lanczos coefficients. Their near overlap over much of the
chain already indicates that the memory-core distinction cannot be reduced
to a simple difference in the overall Lanczos envelope.

\begin{figure}[h!]
\centering
\includegraphics[width=0.35\textwidth]{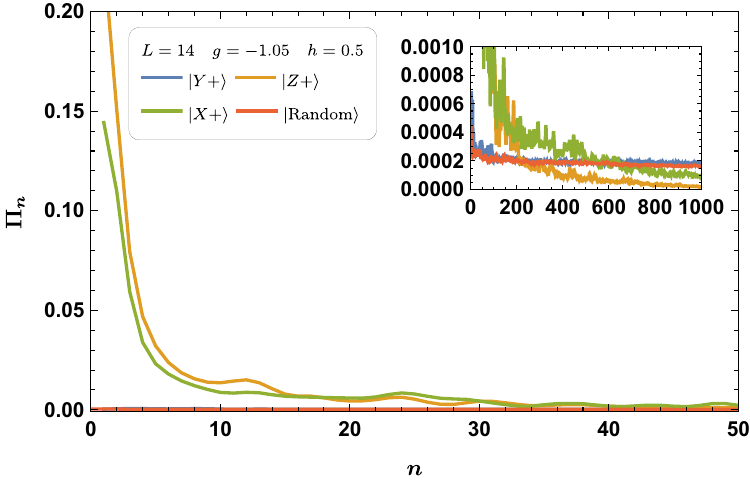}
\caption{
Stationary Krylov occupation profile $\Pi_n$ for the chaotic mixed-field
Ising chain. The structured states exhibit strong low-depth peaks together
with broader stationary tails. The profile specifies where late-time
probability resides; it does not by itself identify the active memory core.
The inset extends the comparison to larger Krylov depths.
}
\label{fig:ising_pi}
\end{figure}

\paragraph{Stationary occupation background.}
The stationary profiles in Fig.~\ref{fig:ising_pi} display pronounced
low-depth peaks for the structured states, but these peaks coexist with
tails extending to much larger depths. The cumulative data make this
distinction quantitative: for $|X+\rangle$ and $|Z+\rangle$,
$n_\Pi^{(0.95)}=2053$ and $416$, respectively, even though their active
memory scales will be only a few tens of Krylov steps. Thus $\Pi_n$
provides the stationary probability background but does not determine the
memory-core boundary.

\begin{figure}[h!]
\centering
\includegraphics[width=0.35\textwidth]{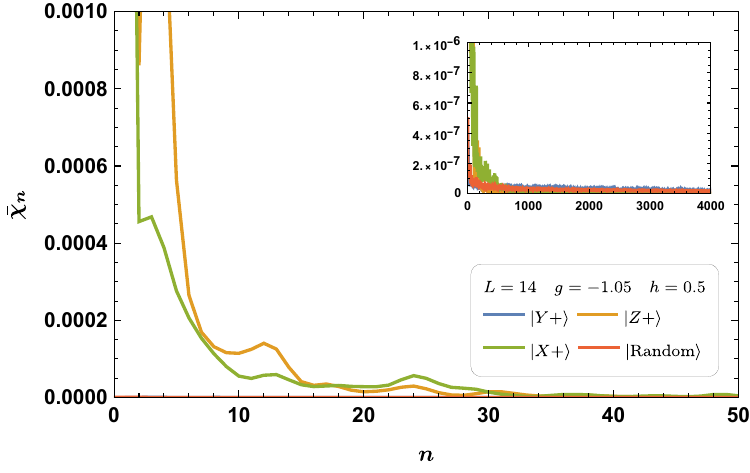}
\caption{
Residual equilibration-fluctuation profile $\overline{\chi_n}$ for the
chaotic mixed-field Ising chain. The states $|X+\rangle$ and $|Z+\rangle$
carry appreciable low-depth fluctuation weight. The efficiently
thermalizing state $|Y+\rangle$ and the random controls have much weaker
signals whose cumulative weight is distributed over a large fraction of
the Krylov chain. The inset shows the larger-depth behavior.
}
\label{fig:ising_chi}
\end{figure}

\begin{figure}[h!]
\centering
\includegraphics[width=0.35\textwidth]{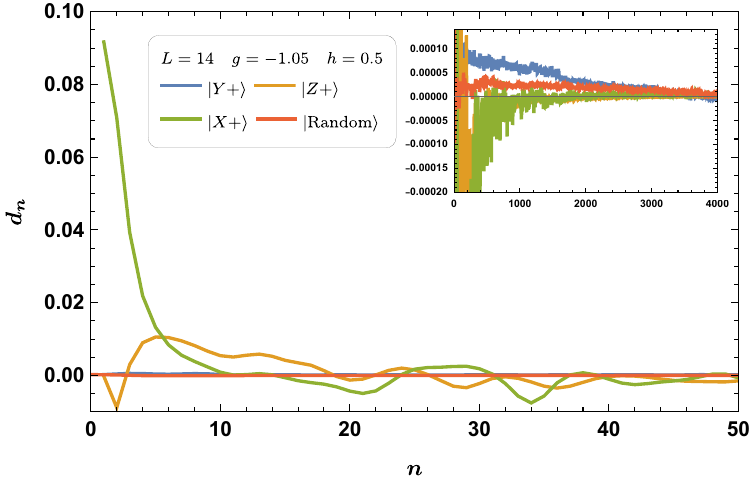}
\caption{
Signed Gibbs-deviation profile
$d_n=\Pi_n-\Pi_n^{(\mathrm{Gibbs})}$ for the chaotic mixed-field Ising
chain. The weakly thermalizing state $|X+\rangle$ exhibits the largest
low-depth Gibbs-reference mismatch. The corresponding mismatch for
$|Z+\rangle$ is compact but considerably weaker, while $|Y+\rangle$ and
the random controls remain close to the Gibbs benchmark.
}
\label{fig:ising_d}
\end{figure}

\begin{figure}[h!]
\centering
\includegraphics[width=0.35\textwidth]{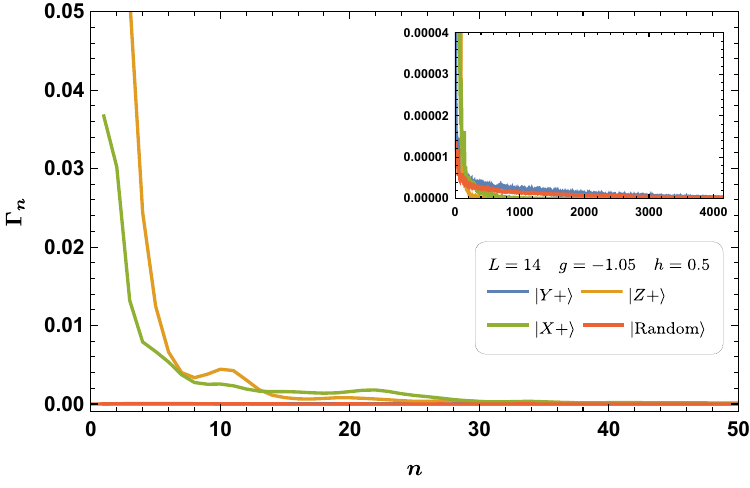}
\caption{
Krylov-current fluctuation activity $\Gamma_n$ for the chaotic
mixed-field Ising chain. The states $|X+\rangle$ and $|Z+\rangle$ support
strong back-and-forth probability-current fluctuations near the Krylov
origin. The efficiently thermalizing and random controls have much weaker
or broadly distributed activity. The inset shows the extended-depth tail.
}
\label{fig:ising_gamma}
\end{figure}

\paragraph{Local active diagnostics.}
Figures~\ref{fig:ising_chi}--\ref{fig:ising_gamma} show the three residual
profiles. The fluctuation and current-activity signals are compact for
both $|X+\rangle$ and $|Z+\rangle$, whereas the Gibbs-reference mismatch
most clearly separates their physical character. The weakly thermalizing
state $|X+\rangle$ has the dominant low-depth Gibbs-deviation weight.
The state $|Z+\rangle$ instead carries stronger total fluctuation and
current activity but a substantially smaller Gibbs mismatch. The
efficiently thermalizing and random controls have weak residual signals
and broad characteristic depths.

\begin{table*}[t]
\centering
\tiny
\setlength{\tabcolsep}{2.5pt}
\resizebox{\textwidth}{!}{
\begin{tabular}{c c c c c c c c c c}
\hline\hline
State
&
$\mathcal D_0$
&
$\mathcal X$
&
$\mathcal N$
&
$\mathcal G$
&
$n_\Pi^{(0.95)}$
&
$n_\chi^{(0.95)}$
&
$n_d^{(0.95)}$
&
$n_\Gamma^{(0.95)}$
&
$n_{\rm mc}$
\\
\hline
$|X+\rangle$
& $8252$
& $6.52\times10^{-3}$
& $1.62\times10^{-2}$
& $1.46\times10^{-1}$
& $2053$
& $25$
& $3$
& $32$
& $32$
\\
$|Z+\rangle$
& $8221$
& $1.88\times10^{-2}$
& $8.92\times10^{-4}$
& $4.28\times10^{-1}$
& $416$
& $10$
& $36$
& $11$
& $36$
\\
$|Y+\rangle$
& $8256$
& $1.34\times10^{-4}$
& $1.86\times10^{-5}$
& $7.67\times10^{-3}$
& $7386$
& $6668$
& $8111$
& $4134$
& $8111$
\\
Random
& $8256$
& $3.40\times10^{-5}$
& $3.10\times10^{-5}$
& $1.33\times10^{-3}$
& $7475$
& $6867$
& $7968$
& $4603$
& $7968$
\\
\hline\hline
\end{tabular}}
\caption{
Integrated residual weights and $95\%$ cumulative depths for the chaotic
mixed-field Ising chain. The entry ``Random'' denotes the ensemble average
over ten symmetry-matched random product controls. The weakly
thermalizing state $|X+\rangle$ has the strongest non-Gibbs core:
$n_{\rm mc}=32$ while the stationary occupation extends to
$n_\Pi^{(0.95)}=2053$. The intermediate state $|Z+\rangle$ also has a
compact active region, $n_{\rm mc}=36$, but its physical composition is
different: it is fluctuation- and activity-dominated and has a much smaller
Gibbs-deviation weight. The efficiently thermalizing and random controls
have tiny residual weights and active depths comparable to the full cyclic
space.
}
\label{tab:ising_total_weights}
\end{table*}

Table~\ref{tab:ising_total_weights} separates the strength and spatial
extent of the stationary residual structure. For $|X+\rangle$,
$n_{\rm mc}/\mathcal D_0\simeq3.9\times10^{-3}$ and the occupation depth is
about sixty-four times larger than the core depth. The state $|Z+\rangle$
has a similarly small active fraction,
$n_{\rm mc}/\mathcal D_0\simeq4.4\times10^{-3}$, but its
Gibbs-deviation weight is about eighteen times smaller than that of
$|X+\rangle$. It therefore represents an intermediate, activity-dominated
memory core rather than the strongly non-Gibbs core of the weakly
thermalizing state. By contrast, $n_{\rm mc}$ occupies almost the entire
cyclic space for $|Y+\rangle$ and the random controls, and their total
residual weights are correspondingly weak. The comparison under the same
Hamiltonian demonstrates that a pronounced memory core is state selective,
not a generic consequence of chaotic dynamics or of the Krylov
construction.

\subsection{Confinement in the mixed-field Ising chain}
\label{subsec:confining_ising_num}

We next use Eq.~\eqref{eq:ising_hamiltonian_num} in the confinement regime
\begin{equation}
J=1,
\qquad
g=0.5,
\qquad
h=0.1,
\qquad
L=14.
\label{eq:confinement_parameters}
\end{equation}
The longitudinal field lifts the degeneracy of the two ferromagnetic
vacua. Domains of the unfavored orientation then form false-vacuum strings
whose energy grows with their length, producing a confining potential
between domain walls
\cite{McCoyWu1978,Kormos2017Confinement,James2019Confinement}.

We compare four reflection-even initial states. In the bit-string
convention $0\equiv|Z+\rangle$ and $1\equiv|Z-\rangle$, these are the
reflection-symmetrized domain-wall state
$|00000001111111\rangle$, denoted $|{\rm DW}\rangle_R$; the bubble state
$|00001111110000\rangle$, denoted $|{\rm B}\rangle_R$; the
reflection-even N\'eel state $|{\rm N}\rangle_R$ obtained from the two
alternating configurations; and a reflection-even random product control
$|{\rm Rand}\rangle_R$. The common reflection sector has dimension
$8256$, while the exact cyclic dimensions are listed in
Table~\ref{tab:confinement_total_weights}.

\begin{figure}[t]
\centering
\includegraphics[width=0.35\textwidth]{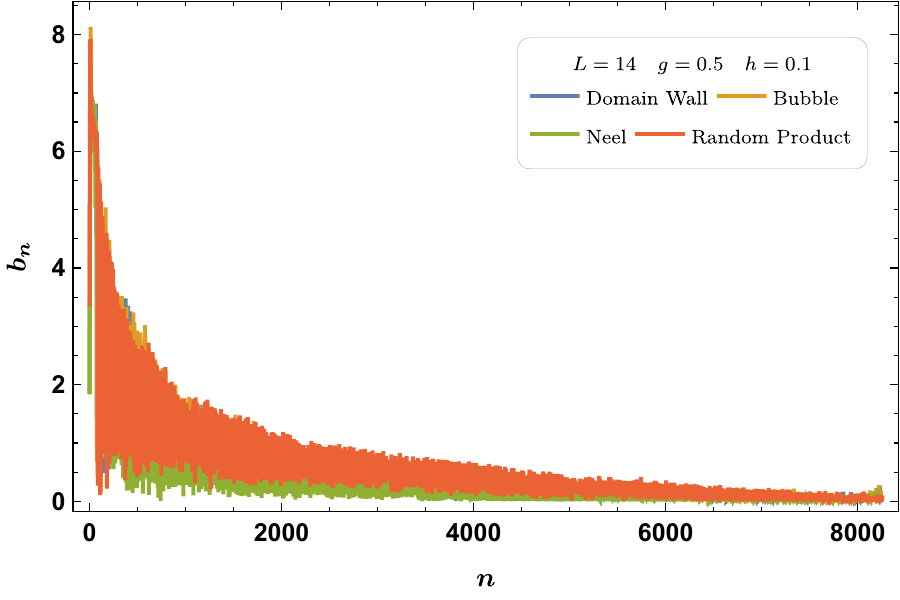}
\caption{
Lanczos coefficients $b_n$ in the confinement regime
($L=14$, $J=1$, $g=0.5$, $h=0.1$). The domain-wall, bubble, N\'eel, and
random states all explore cyclic spaces close to the full reflection-even
sector, but their early-depth Lanczos structures differ. Exact cyclic
dimensions are reported in Table~\ref{tab:confinement_total_weights}.
}
\label{fig:confinementbn}
\end{figure}

Figure~\ref{fig:confinementbn} shows that the three structured states have
different early Lanczos patterns, consistent with their distinct
domain-wall content. At large depth, however, all four chains explore
nearly the entire reflection-even sector. The memory distinction must
therefore come from the stationary profiles rather than from a small
accessible Hilbert space.

\begin{figure}[h!]
\centering
\includegraphics[width=0.35\textwidth]{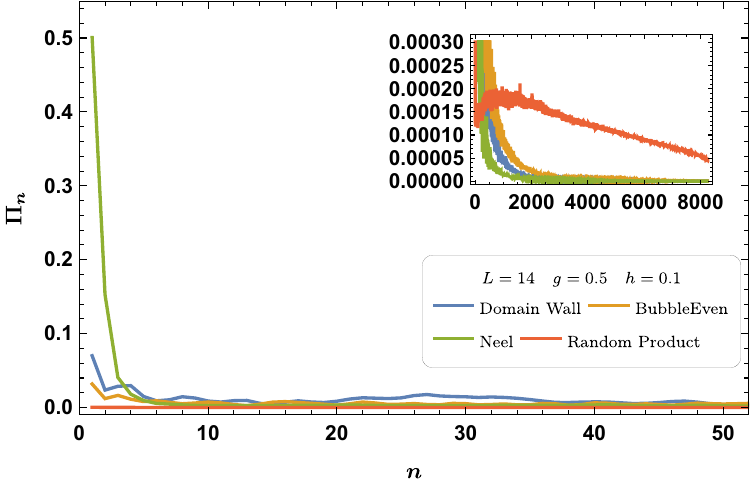}
\caption{
Stationary Krylov occupation profile $\Pi_n$ in the confinement regime.
The confinement-sensitive states have pronounced low-depth peaks together
with extended stationary tails. The random control is broadly distributed.
The occupation profile supplies the stationary background; its cumulative
extent is distinct from the active memory-core depth.
}
\label{fig:confinement_pi}
\end{figure}

\paragraph{Stationary occupation background.}
The profiles in Fig.~\ref{fig:confinement_pi} again show that low-depth
occupation peaks do not exhaust the stationary probability distribution.
For the N\'eel, domain-wall, and bubble states,
$n_\Pi^{(0.95)}=228$, $735$, and $1159$, respectively. These scales are
all substantially larger than the active core depths extracted below.
The stationary occupation therefore contains a broad tail even when the
memory-related residual structure is shallow.

\begin{figure}[h!]
\centering
\includegraphics[width=0.35\textwidth]{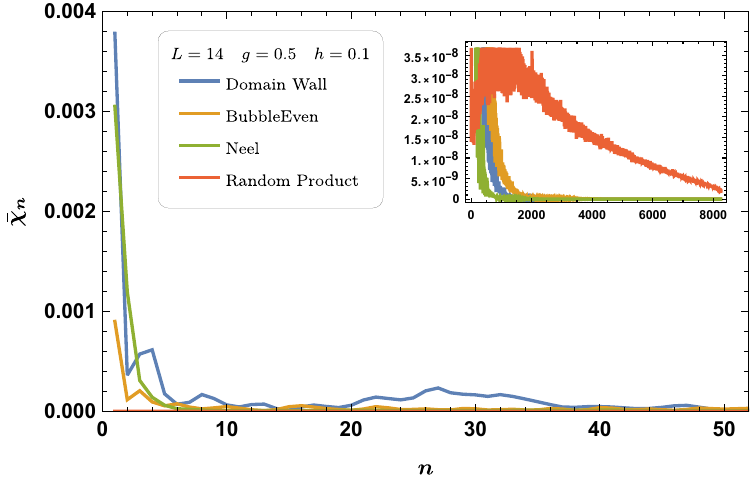}
\caption{
Residual equilibration-fluctuation profile $\overline{\chi_n}$ in the
confinement regime. The N\'eel state has the sharpest low-depth
fluctuation structure; the domain-wall and bubble states retain broader
confinement-sensitive signals. The random control has a much weaker
profile spread over the Krylov chain.
}
\label{fig:confinement_chi}
\end{figure}

\begin{figure}[h!]
\centering
\includegraphics[width=0.35\textwidth]{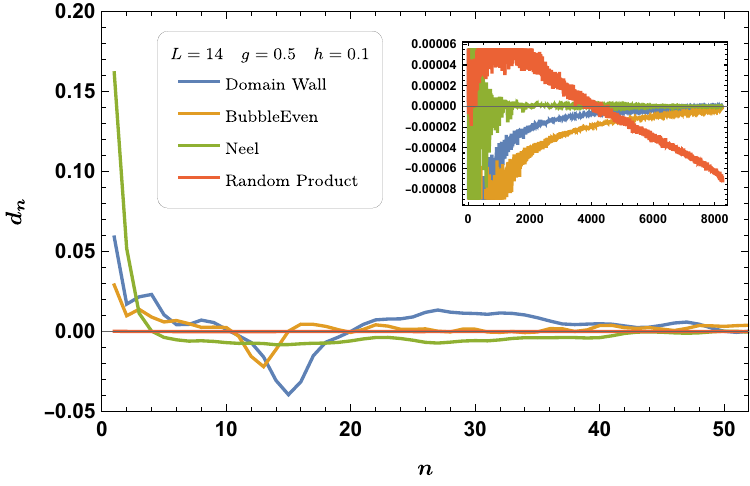}
\caption{
Signed Gibbs-deviation profile
$d_n=\Pi_n-\Pi_n^{(\mathrm{Gibbs})}$ in the confinement regime. The
N\'eel state has the largest and most compact mismatch, while the
domain-wall and bubble states show substantial but broader deviations.
The random control remains close to the Gibbs benchmark.
}
\label{fig:confinement_d}
\end{figure}

\begin{figure}[h!]
\centering
\includegraphics[width=0.35\textwidth]{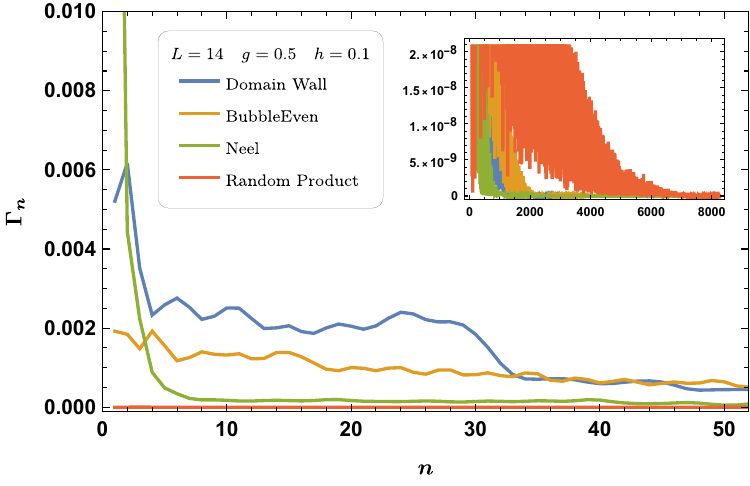}
\caption{
Krylov-current fluctuation activity $\Gamma_n$ in the confinement
regime. The N\'eel state supports a sharp low-depth activity profile.
The domain-wall and bubble states remain dynamically active over broader
intervals, whereas the random control has a much weaker signal.
}
\label{fig:confinement_gamma}
\end{figure}

\paragraph{Local active diagnostics.}
Figures~\ref{fig:confinement_chi}--\ref{fig:confinement_gamma} reveal
three related but nonidentical core morphologies. The N\'eel state has the
most compact and largest Gibbs-deviation signal. The domain-wall state has
a strong, balanced set of fluctuation, mismatch, and current-activity
weights over a slightly broader interval. The bubble state remains
clearly non-generic but distributes its residual content over a wider
low-depth window. The random control has weak total weights and no
compact active interval.

\begin{table*}[t]
\centering
\tiny
\setlength{\tabcolsep}{2.5pt}
\resizebox{\textwidth}{!}{
\begin{tabular}{c c c c c c c c c c}
\hline\hline
State
&
$\mathcal D_0$
&
$\mathcal X$
&
$\mathcal N$
&
$\mathcal G$
&
$n_\Pi^{(0.95)}$
&
$n_\chi^{(0.95)}$
&
$n_d^{(0.95)}$
&
$n_\Gamma^{(0.95)}$
&
$n_{\rm mc}$
\\
\hline
$|{\rm N}\rangle_R$
& $8226$
& $5.20\times10^{-3}$
& $3.02\times10^{-2}$
& $4.42\times10^{-2}$
& $228$
& $26$
& $1$
& $41$
& $41$
\\
$|{\rm DW}\rangle_R$
& $8240$
& $1.00\times10^{-2}$
& $1.14\times10^{-2}$
& $9.56\times10^{-2}$
& $735$
& $57$
& $41$
& $56$
& $57$
\\
$|{\rm B}\rangle_R$
& $8244$
& $3.91\times10^{-3}$
& $2.96\times10^{-3}$
& $7.73\times10^{-2}$
& $1159$
& $144$
& $116$
& $129$
& $144$
\\
$|{\rm Rand}\rangle_R$
& $8256$
& $1.31\times10^{-4}$
& $8.96\times10^{-6}$
& $4.34\times10^{-4}$
& $7392$
& $6851$
& $8133$
& $2296$
& $8133$
\\
\hline\hline
\end{tabular}}
\caption{
Integrated residual weights and $95\%$ cumulative depths in the
confinement regime. 
The reflection-even N\'eel
state has the most compact
and most strongly non-Gibbs core, $n_{\rm mc}=41$. The domain-wall state
also has a pronounced compact core, $n_{\rm mc}=57$, while the bubble
state has a broader confinement-induced core, $n_{\rm mc}=144$. In all
three cases the occupation depth is several times larger than the active
depth, revealing a stationary halo. The random control has tiny residual
weights and an active depth comparable to the full cyclic space.
}
\label{tab:confinement_total_weights}
\end{table*}

The three confinement-sensitive states therefore exhibit a hierarchy of
memory-core structures rather than a single identical profile. Their active
fractions are small:
$n_{\rm mc}/\mathcal D_0\simeq5.0\times10^{-3}$ for the N\'eel state,
$6.9\times10^{-3}$ for the domain wall, and
$1.8\times10^{-2}$ for the bubble. Their occupation depths are about six,
thirteen, and nine times larger than their respective core depths. The
random control behaves oppositely: its total weights are orders of
magnitude smaller, and its largest active threshold reaches almost the full
cyclic dimension. Confinement therefore produces state-selective active
cores with different physical compositions and spatial extents.


\subsection{PXP model and quantum many-body scars}
\label{subsec:pxp_num}

We finally ask whether the same Krylov-space organization appears in a constrained model where anomalous relaxation arises from quantum many-body scars. We study the periodic PXP Hamiltonian~\cite{Lesanovsky:2012kzj}
\begin{equation}
    H_{\rm PXP}
    =
    \sum_{i=1}^{L} P_{i-1}X_iP_{i+1},
    \label{eq:pxp_hamiltonian_num}
\end{equation}
where
\begin{equation}
    P_i={1+\sigma_i^z\over 2},
    \qquad
    n_i={1-\sigma_i^z\over 2},
\end{equation}
so that $P_i=1-n_i$. The projectors enforce the Rydberg blockade constraint $n_i n_{i+1}=0$, which forbids simultaneous excitation of neighboring sites.

The PXP model is nonintegrable and thermalizes for generic initial states, but it contains atypical scarred eigenstates embedded in an otherwise thermal spectrum~\cite{Bernien:2017ubn, turner2018quantum,turner2018weak,shiraishi2017systematic, mori2017thermalization,Moudgalya:2021xlu}. These states produce long-lived oscillations and anomalously slow relaxation for special initial states, most prominently the N\'eel state $|Z_2\rangle=|101010\ldots\rangle$.

We consider a periodic chain of length
$L=28$ and compare the scarred N\'eel state with a non-scarred reference state,
\begin{align}
\label{eq:neel}
|Z_2\rangle
&=
|101010\cdots 10\rangle ,
\\
|G\rangle
&=
\frac{|Z_4\rangle+T^2|Z_4\rangle}{\sqrt{2}},
\qquad
|Z_4\rangle=|10001000\cdots1000\rangle ,
\nonumber
\end{align}
where $T$ is the one-site translation operator. Both states satisfy $T^2|\psi\rangle=|\psi\rangle$, so they are analyzed in the same translation setting. The state $|Z_2\rangle$ has strong overlap with the scarred eigenstates, while $|G\rangle$ is a non-scarred reference state with compatible translation structure.

Krylov dynamics and spread complexity for scarred PXP initial states have been studied in Refs.~\cite{PhysRevB.106.205150,Nandy_2024, Caputa:2025ucl}. A characteristic feature of the scarred N\'eel state is that the first $O(L)$ Lanczos coefficients approximately form an arch, reminiscent of an emergent $\mathrm{SU}(2)$ structure~\cite{Caputa:2021sib,Choi:2018cfo}. The PXP model, however, does not realize an exact finite-dimensional $\mathrm{SU}(2)$ representation. The Lanczos coefficients do not terminate at $n=L$; they continue beyond this scale and eventually depart from the approximate arch. In this work we do not impose a forward-scattering truncation. All profiles below are computed in the full numerically resolved cyclic Krylov space generated by the PXP Hamiltonian.

\begin{figure}[t]
\centering
\includegraphics[width=0.36\textwidth]{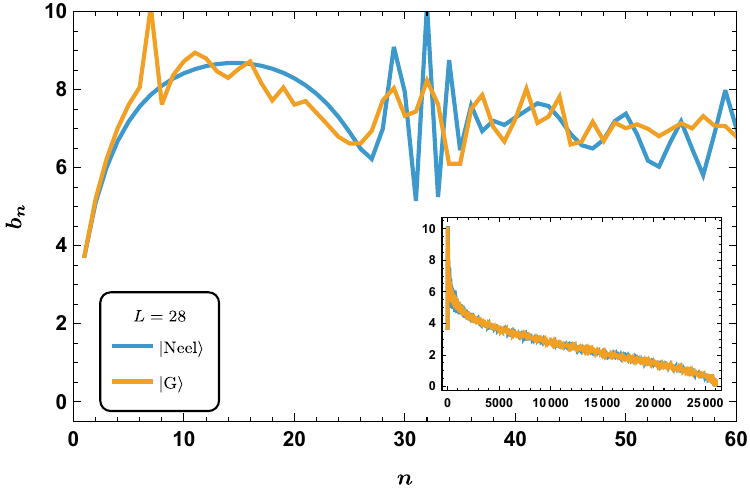}
\caption{
Lanczos coefficients $b_n$ for the periodic $L=28$ PXP model, constructed
from the degeneracy-resolved active spectral measures of the scarred
N\'eel state and the non-scarred reference. The cyclic dimensions used in
the numerical profiles are $\mathcal D_0(|Z_2\rangle)=26015$ and
$\mathcal D_0(|G\rangle)=26021$. The scar-related $O(L)$ arch occupies only the beginning of the much longer numerically resolved cyclic chains.
}
\label{fig:pxpbn}
\end{figure}

The full periodic constrained Hilbert
space has dimension $710647$.
At the energy-grouping and spectral-weight tolerances specified in
Appendix~\ref{app:pxp_dimension_count}, the active-spectral construction
yields the numerically resolved cyclic dimensions
\begin{equation}
\mathcal D_0(|Z_2\rangle)=26015,
\qquad
\mathcal D_0(|G\rangle)=26021 .
\end{equation}
These dimensions refer to the full numerically resolved cyclic spaces and
not to forward-scattering truncations.
Figure~\ref{fig:pxpbn} displays the
corresponding Lanczos coefficients.

For the PXP calculations, all commuting spatial symmetries are resolved
before the active spectral measure is constructed.  Numerically degenerate
energy levels are combined using the energy-resolution tolerance
$\varepsilon_E$, groups with spectral weight below $\varepsilon_\omega$
are discarded, and the retained active measure is then symmetrized under
the exact chiral pairing $E\leftrightarrow -E$.  At the large active
dimensions used for the production calculations, explicitly enumerating
and sorting all $\mathcal D_0(\mathcal D_0-1)$ ordered nonzero transitions
is impractical.  We therefore evaluate $\overline{\chi_n}$ and $\Gamma_n$
with the chiral-pair-resolved formulas derived in Appendix~A.  These formulas include exactly the
repeated gaps enforced by chiral symmetry, including the transition pairs
that involve the active zero-energy component.

At the validation sizes $L=18,20,22,24$, the scalable chiral calculation
is checked against an independent orbit-compressed complete equal-gap
evaluation. In this validation branch, distinct chiral gap orbits whose
numerical gaps coincide within $\varepsilon_\Delta$ are combined coherently
before the absolute square is taken, and $\varepsilon_\Delta$ is varied over
a logarithmic grid. The comparison monitors the full $\overline{\chi_n}$ and
$\Gamma_n$ profiles and their $95\%$ and $99\%$ cumulative depths. The
production calculation is accepted only when the chiral and complete-gap
results agree within the stated numerical accuracy throughout the sampled
stable $\varepsilon_\Delta$ intervals. Consequently, $\varepsilon_\Delta$ is
a validation parameter; it is not used in the large-$\mathcal D_0$
chiral-pair-resolved production profiles.

\begin{figure}[h!]
\centering
\includegraphics[width=0.36\textwidth]{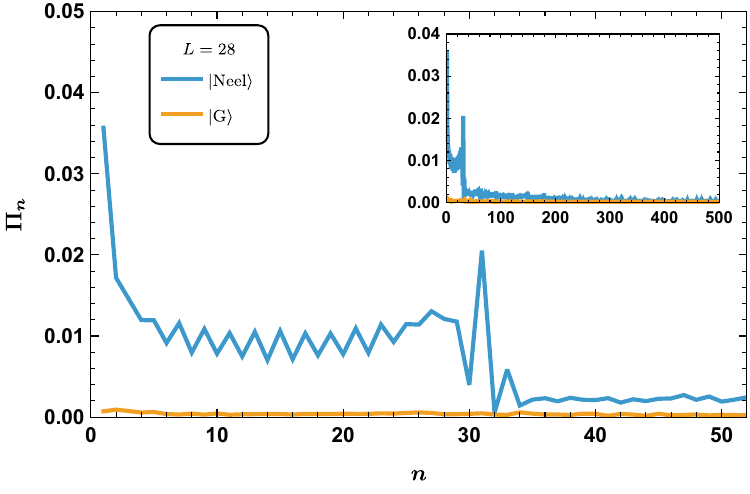}
\caption{
Stationary Krylov occupation profile $\Pi_n$ for the periodic $L=28$ PXP
model. The scarred N\'eel state has a pronounced low-depth peak and a
feature near the scar-related scale, but its stationary probability also
extends through a long tail. The reference state is broader. The inset
shows the larger-depth behavior.
}
\label{fig:pxp_pi}
\end{figure}

\paragraph{Stationary occupation background.}
Figure~\ref{fig:pxp_pi} shows a strong low-depth concentration for
$|Z_2\rangle$, including the scar-related structure near $n\simeq L$.
The cumulative occupation, however, reaches $95\%$ only at
$n_\Pi^{(0.95)}=9818$. The low-depth peak is therefore embedded in an
extensive stationary tail and cannot by itself define the memory core.
The reference state is broader still, with
$n_\Pi^{(0.95)}=21299$.

\begin{figure}[h!]
\centering
\includegraphics[width=0.36\textwidth]{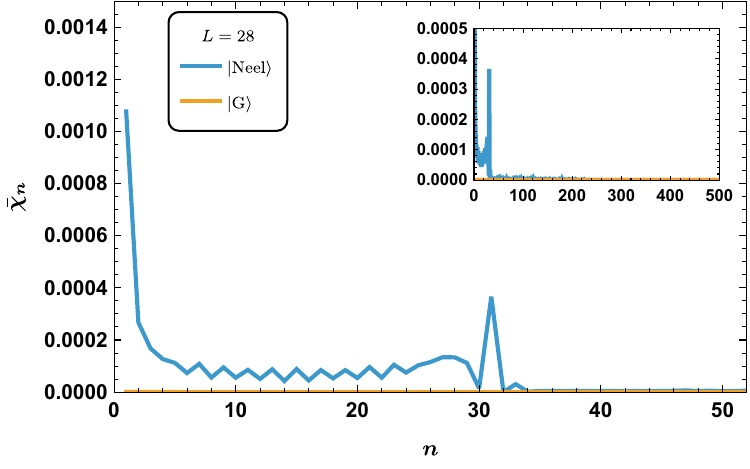}
\caption{
Residual equilibration-fluctuation profile $\overline{\chi_n}$ for the
periodic $L=28$ PXP model. The scarred N\'eel state has an appreciable
low-depth signal with a feature near $n\simeq L$, whereas the reference
state has a much weaker profile.
}
\label{fig:pxp_chi}
\end{figure}

\begin{figure}[h!]
\centering
\includegraphics[width=0.36\textwidth]{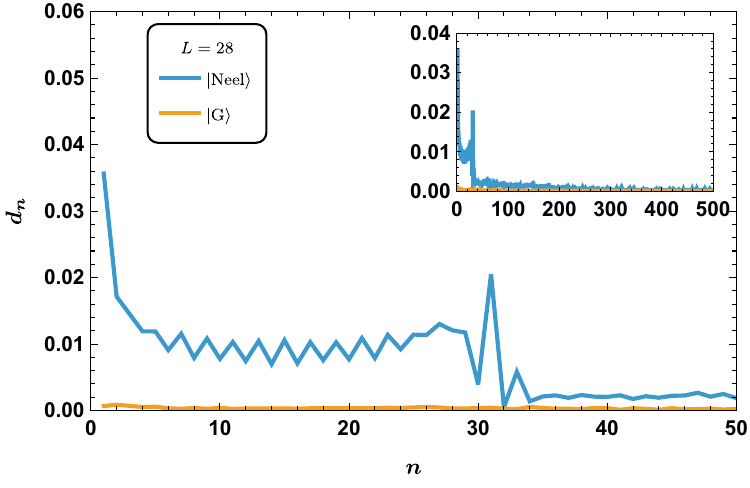}
\caption{
Signed Gibbs-deviation profile
$d_n=\Pi_n-\Pi_n^{(\mathrm{Gibbs})}$ for the periodic $L=28$ PXP model.
The scarred N\'eel state has a pronounced low-depth mismatch, including
the scar-related feature near $n\simeq L$. The non-scarred reference
remains much closer to the Gibbs benchmark.
}
\label{fig:pxp_d}
\end{figure}

\begin{figure}[h!]
\centering
\includegraphics[width=0.36\textwidth]{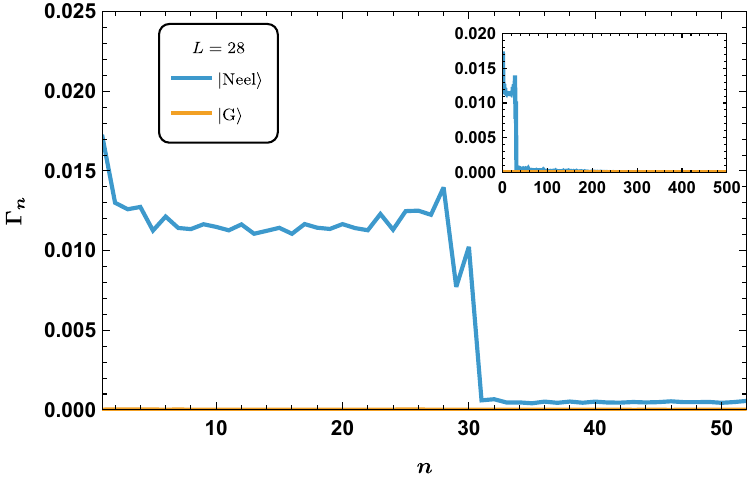}
\caption{
Krylov-current fluctuation activity $\Gamma_n$ for the periodic $L=28$
PXP model. The scarred state supports strong back-and-forth current
fluctuations within the low-depth scar-related region, whereas the
reference activity is much weaker.
}
\label{fig:pxp_gamma}
\end{figure}

\paragraph{Local active diagnostics.}
Figures~\ref{fig:pxp_chi}--\ref{fig:pxp_gamma} show that the scarred state
carries all three residual signals in the first few hundred Krylov steps,
with prominent structure near $n\simeq L$. The reference state has
residual weights smaller by approximately two orders of magnitude and
accumulates those weak weights over most of the numerically resolved cyclic space.

\begin{table*}[t]
\centering
\tiny
\setlength{\tabcolsep}{2.5pt}
\resizebox{\textwidth}{!}{
\begin{tabular}{c c c c c c c c c c}
\hline\hline
State
&
$\mathcal D_0$
&
$\mathcal X$
&
$\mathcal N$
&
$\mathcal G$
&
$n_\Pi^{(0.95)}$
&
$n_\chi^{(0.95)}$
&
$n_d^{(0.95)}$
&
$n_\Gamma^{(0.95)}$
&
$n_{\rm mc}$
\\
\hline
$|Z_2\rangle$
& $26015$
& $4.68\times10^{-3}$
& $5.45\times10^{-3}$
& $3.97\times10^{-1}$
& $9818$
& $109$
& $100$
& $92$
& $109$
\\
$|G\rangle$
& $26021$
& $7.34\times10^{-5}$
& $3.58\times10^{-5}$
& $3.75\times10^{-3}$
& $21299$
& $14541$
& $24247$
& $6346$
& $24247$
\\
\hline\hline
\end{tabular}}
\caption{
Integrated residual weights and $95\%$ cumulative depths for the periodic
$L=28$ PXP model. The scarred N\'eel state has
$n_{\rm mc}=109$, only about $0.42\%$ of its numerically resolved cyclic dimension, while
its stationary occupation extends to $n_\Pi^{(0.95)}=9818$. The reference
state has residual weights smaller by roughly two orders of magnitude and
$n_{\rm mc}=24247$, comparable to its full cyclic dimension. The contrast
demonstrates a compact scarred memory core embedded in a broad stationary
halo.
}
\label{tab:pxp_total_weights}
\end{table*}

Table~\ref{tab:pxp_total_weights} gives the sharpest core--halo
separation in the paper. For $|Z_2\rangle$, the occupation depth is about
ninety times larger than $n_{\rm mc}$, while the active core occupies only
$4.2\times10^{-3}$ of the cyclic space. Relative to $|G\rangle$, the
scarred state has approximately sixty-four times larger $\mathcal X$,
one hundred fifty times larger $\mathcal N$, and one hundred six times
larger $\mathcal G$. The reference has neither appreciable residual
weights nor a compact common active interval.

\subsection{Cumulative comparison and the core--halo structure}
\label{subsec:cumulative_core}

\begin{figure}[!t]
\centering
\includegraphics[width=0.3\textwidth]{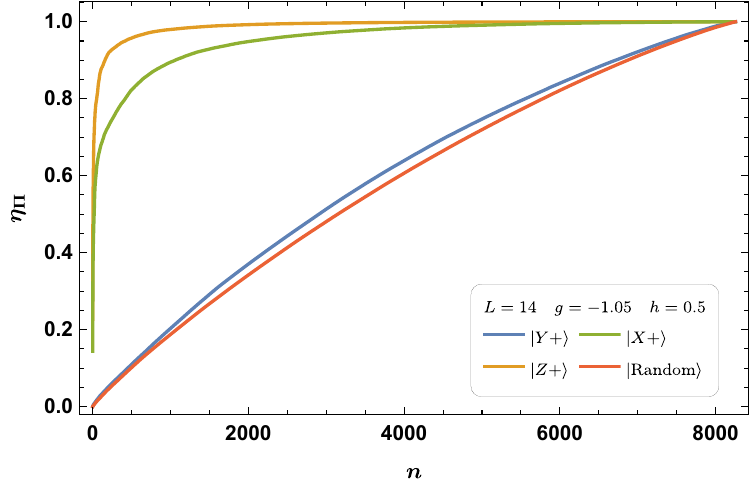}\\[0.5em]
\includegraphics[width=0.3\textwidth]{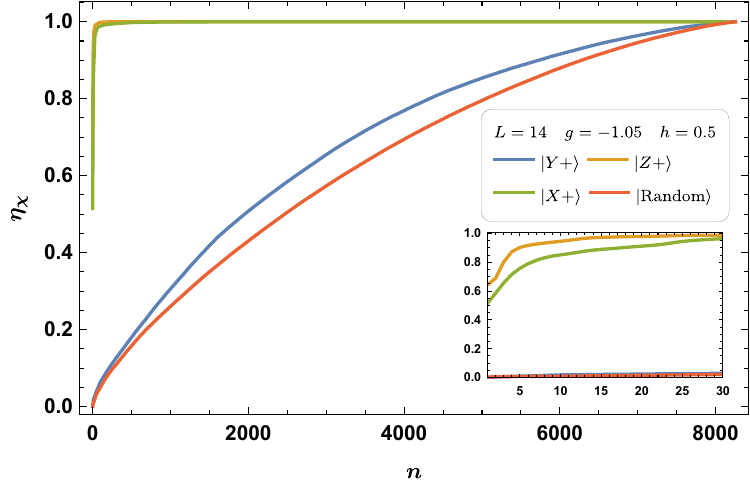}\\[0.5em]
\includegraphics[width=0.3\textwidth]{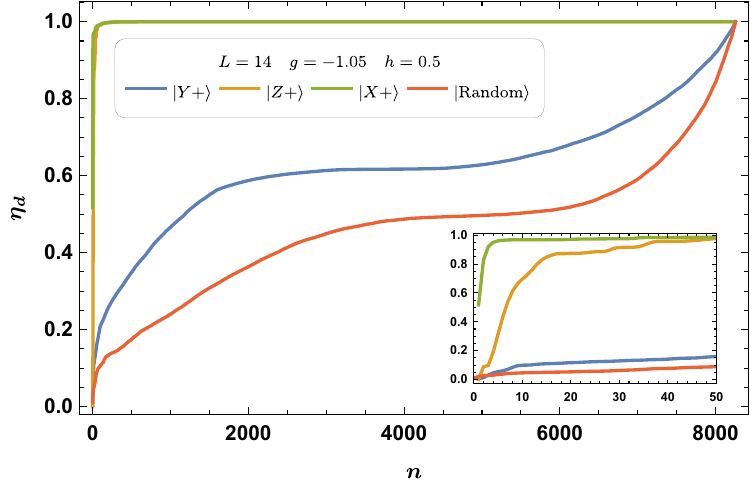}\\[0.5em]
\includegraphics[width=0.3\textwidth]{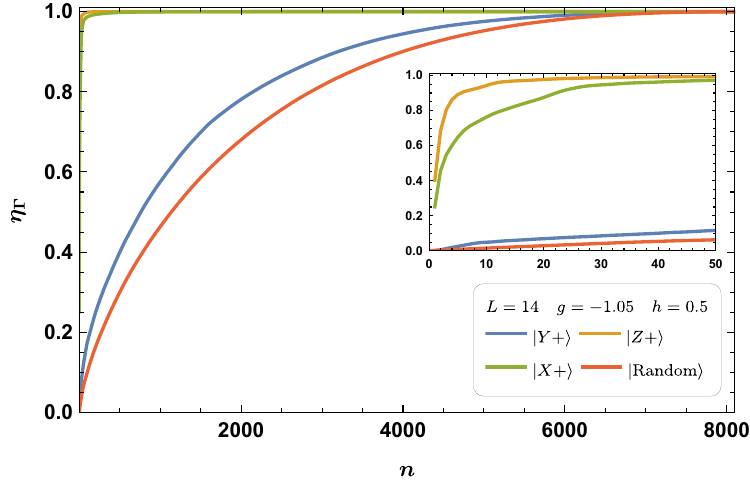}
\caption{
Cumulative profiles for the chaotic mixed-field Ising chain. From top to
bottom: $\eta_\Pi$, $\eta_\chi$, $\eta_d$, and $\eta_\Gamma$. The active
profiles of $|X+\rangle$ and $|Z+\rangle$ saturate within
$n_{\rm mc}=32$ and $36$, respectively, while their occupation clouds
extend to $n_\Pi^{(0.95)}=2053$ and $416$. This separation reveals compact
active cores inside broader stationary halos. The efficiently
thermalizing and random controls have weak residual weights and active
depths comparable to the full cyclic space.
}
\label{fig:cumulative_cores}
\end{figure}

\begin{figure}[!t]
\centering
\includegraphics[width=0.3\textwidth]{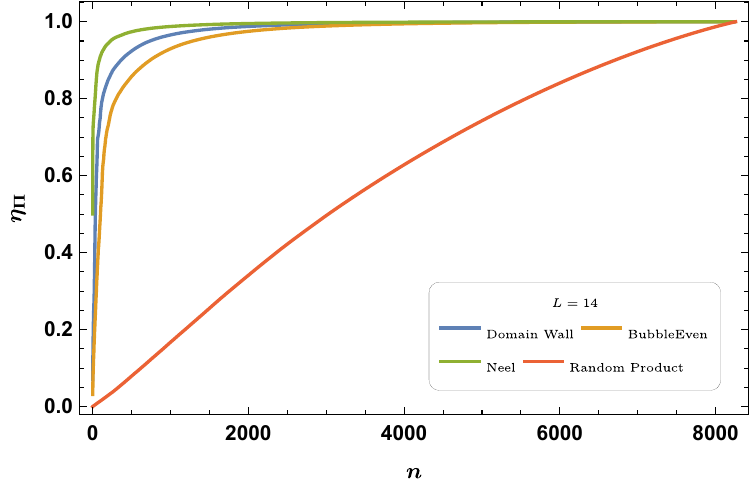}\\[0.5em]
\includegraphics[width=0.3\textwidth]{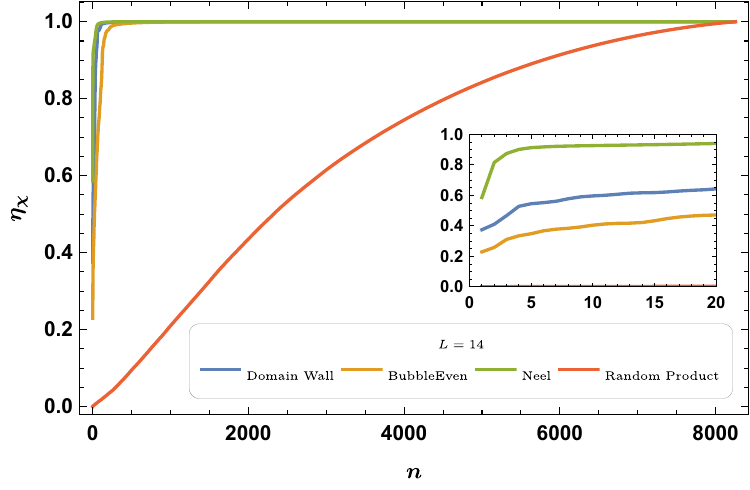}\\[0.5em]
\includegraphics[width=0.3\textwidth]{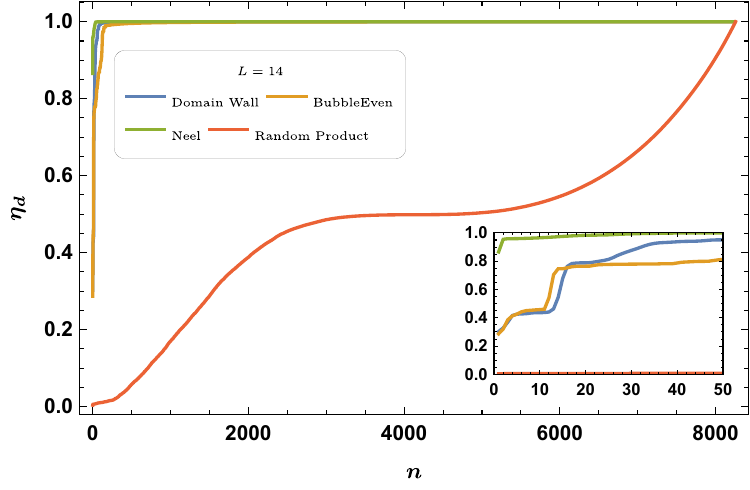}\\[0.5em]
\includegraphics[width=0.3\textwidth]{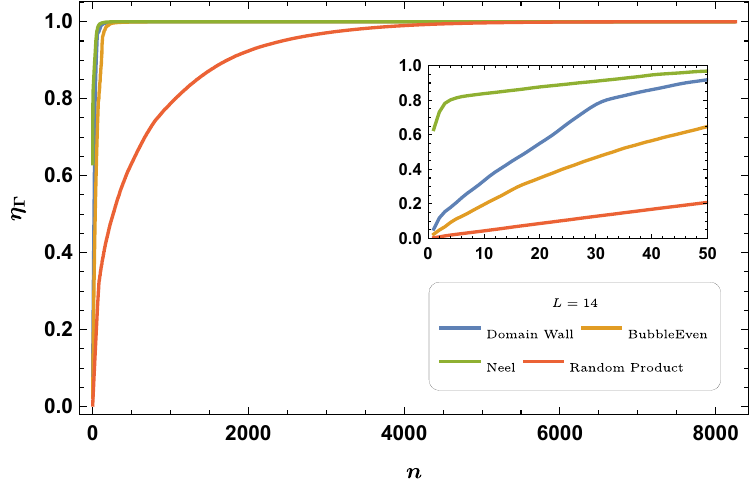}
\caption{
Cumulative profiles in the confinement regime. From top to bottom:
$\eta_\Pi$, $\eta_\chi$, $\eta_d$, and $\eta_\Gamma$. The N\'eel and
domain-wall states have compact active cores, while the bubble state has
a broader but still non-generic core. In each confinement-sensitive state,
the stationary occupation extends substantially beyond the active interval.
The random control has tiny residual weights and broad cumulative profiles.
}
\label{fig:cumulative_cores_confinement}
\end{figure}

\begin{figure}[!t]
\centering
\includegraphics[width=0.3\textwidth]{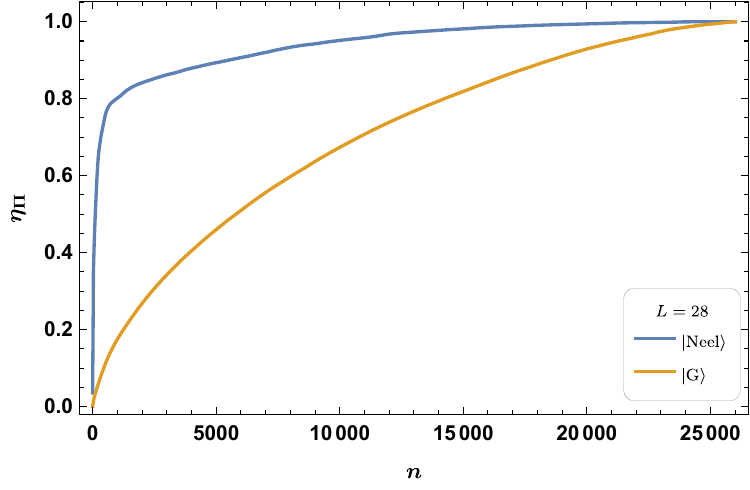}\\[0.5em]
\includegraphics[width=0.3\textwidth]{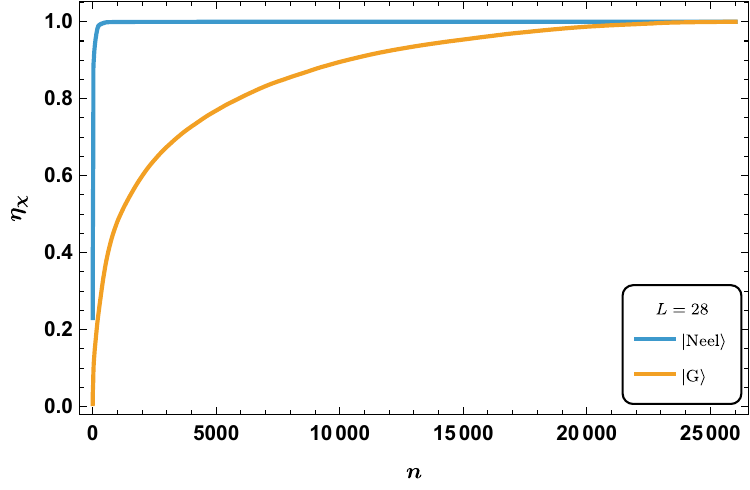}\\[0.5em]
\includegraphics[width=0.3\textwidth]{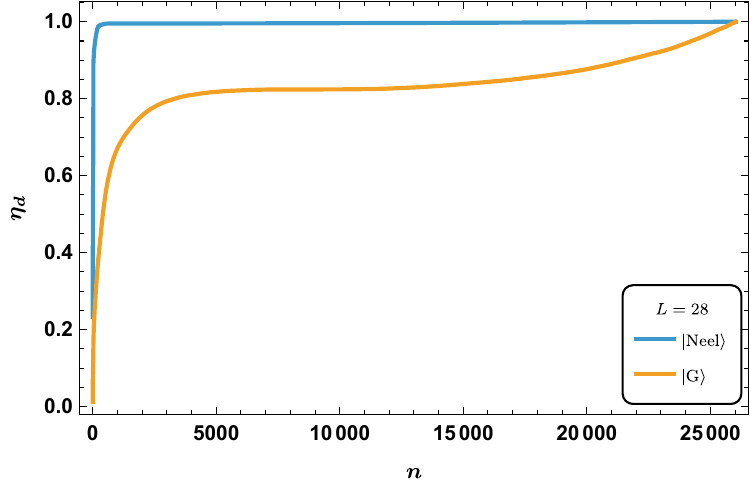}\\[0.5em]
\includegraphics[width=0.3\textwidth]{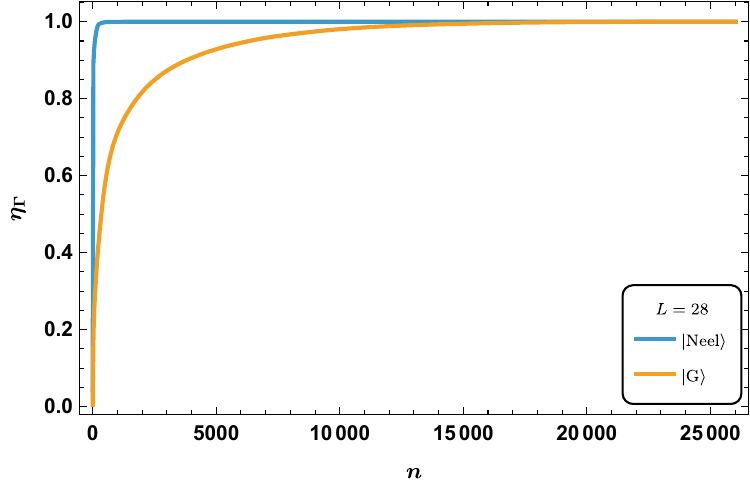}
\caption{
Cumulative profiles for the periodic $L=28$ PXP model. From top to bottom:
$\eta_\Pi$, $\eta_\chi$, $\eta_d$, and $\eta_\Gamma$. For the scarred
N\'eel state, the three active curves saturate by $n_{\rm mc}=109$, whereas
$95\%$ of the stationary occupation requires $9818$ Krylov depths. The
reference state has weak residual weights and accumulates them over nearly
the full cyclic space.
}
\label{fig:cumulative_cores-PXP}
\end{figure}

The cumulative profiles in
Figs.~\ref{fig:cumulative_cores}--\ref{fig:cumulative_cores-PXP}
make the distinction between the stationary probability cloud and the
memory-bearing core explicit. In each figure the upper panel shows
$\eta_\Pi$, while the lower three panels show the normalized active
weights $\eta_\chi$, $\eta_d$, and $\eta_\Gamma$. The memory core is
identified from the lower three curves and their total weights, not from
the occupation curve.

Across the three models, the recurring result is therefore not a uniformly
localized stationary state. It is a compact interval containing
appreciable fluctuation, Gibbs-deviation, and current-activity weights
inside a much broader stationary probability cloud. The occupation
bounds derived in Sec.~\ref{sec:stationary} explain why the active profiles
must lie within the occupied background, but they do not fix their total
weights or their characteristic depths. The observed combination
\begin{equation}
\mathcal X,\mathcal N,\mathcal G\ \text{appreciable},
\qquad
n_{\rm mc}\ll n_\Pi^{(0.95)},
\qquad
n_{\rm mc}\ll\mathcal D_0 .
\label{eq:core_signature}
\end{equation}
is therefore the nontrivial numerical signature of the anomalous states.

Comparable values of
$n_\chi^{(0.95)}$, $n_d^{(0.95)}$, and
$n_\Gamma^{(0.95)}$ define a common enclosing interval; they do not imply
that the three local profiles are pointwise identical. The local figures
show that their dominant features overlap within that interval, while the
different total weights distinguish the physical character of each core.
This distinction is especially clear in the chaotic Ising chain:
$|X+\rangle$ has the strongest non-Gibbs core, whereas $|Z+\rangle$ has a
compact but more activity-dominated core.

The finite-size data in Appendix~\ref{app:core_scaling} extend this
comparison to all three nonintegrable settings. Over the accessible sizes,
the active depth is compatible with growth no faster than approximately
linear in $L$, while the exact cyclic dimension grows much more rapidly.
Equivalently, the data are consistent with
\begin{equation}
n_{\rm mc}=O(L),
\qquad
\frac{n_{\rm mc}}{\mathcal D_0}\longrightarrow0,
\end{equation}
provided the observed finite-size trends persist. This statement is an
extrapolation from finite systems, but it supports the interpretation of
the memory core as a subextensive active structure rather than a finite
fraction of the exact Krylov space.

The results establish a common stationary phenomenology across weak
thermalization, confinement-sensitive dynamics, and many-body scarring.
They do not imply a universal microscopic mechanism, nor do they use the
memory core as a classifier of integrability. The essential comparison is
state selective: special initial states in otherwise thermalizing
nonintegrable systems possess appreciable compact residual structures,
whereas symmetry-matched generic references under the same Hamiltonians do
not.

The next section separates this stationary core--halo structure from Krylov
growth itself by comparing with exactly solvable infinite chains in which
probability escapes to arbitrarily large depth and no normalizable
fixed-depth stationary measure remains.


\FloatBarrier
\section{Memory cores versus Krylov growth}
\label{sec:escaping_geometries}

The numerical results of Sec.~\ref{sec:numerics} show that anomalous
initial states in weak thermalization, confinement-sensitive dynamics, and
many-body scarring develop a common stationary organization in Krylov space:
the residual fluctuation, Gibbs-deviation, and Krylov-current fluctuation
weights become concentrated within a compact low-depth region. This section
addresses a complementary question: whether such a memory core is simply a
consequence of Krylov spreading and large Krylov complexity.

To answer this question, we compare the finite many-body systems studied in
the main text with exactly solvable infinite-dimensional Krylov geometries.
These models exhibit well-characterized spreading laws, including ballistic
propagation, accelerated spreading, and exponential front propagation
associated with asymptotically linear Lanczos growth. However, their
wavepackets escape indefinitely toward large Krylov depth and do not
develop
a normalizable stationary 
fixed-depth Krylov 
distribution~\cite{Alishahiha:2025byi}. 
They therefore
provide a useful null comparison: strong Krylov growth alone does not create
the stationary organization required for a memory core.

For finite-dimensional Krylov spaces, the stationary occupation profile is
defined by taking the infinite-time limit at fixed $\mathcal D_0$,
\begin{equation}
\Pi_n=
\lim_{T\rightarrow\infty}
\frac{1}{T}\int_0^T P_n(t)\,dt ,
\end{equation}
with
\begin{equation}
\sum_{n=0}^{\mathcal D_0-1}\Pi_n=1 .
\end{equation}
The memory-core diagnostics introduced in Sec.~\ref{sec:stationary} are
therefore finite-dimensional stationary quantities.

The distinction becomes important for infinite Krylov chains. The finite-time
average
\begin{equation}
\Pi_n(T)=\frac1T\int_0^T P_n(t)\,dt
\end{equation}
is normalized for every finite $T$, but in an escaping geometry
\begin{equation}
\lim_{T\rightarrow\infty}\Pi_n(T)=0
\end{equation}
for every fixed $n$. Hence
\begin{equation}
\lim_{T\rightarrow\infty}\sum_n\Pi_n(T)=1,
\qquad
\sum_n\lim_{T\rightarrow\infty}\Pi_n(T)=0 .
\end{equation}
The noncommutativity of the infinite-time limit and the infinite Krylov-depth
sum expresses escape: probability is not lost, but it is transferred to
arbitrarily large depths, leaving no normalizable stationary measure.

Although stationary profiles vanish in these escaping systems, transient
integrated quantities can still characterize propagation:
\begin{equation}
\mathcal P_n=\int_0^\infty P_n(t)\,dt,
\;\;\;\;\;\;\;\;
\mathcal Q_n=\int_0^\infty P_n^2(t)\,dt,
\end{equation}
and
\begin{equation}
\mathcal J_n=\int_0^\infty J_n^2(t)\,dt .
\end{equation}
These quantities describe cumulative visitation and current passage during
escape. They are not stationary diagnostics and do not define a memory core.

\subsection{Exactly solvable infinite-dimensional Krylov geometries}

\subsubsection{Constant Lanczos coefficients: ballistic spreading}

We begin with a Krylov chain with constant Lanczos coefficients,
\begin{equation}
a_n = a, \qquad b_n = b, \qquad n \ge 1,
\end{equation}
where the constant diagonal term $a$ only produces an overall phase and may be set to zero. This defines a semi-infinite Jacobi problem with uniform hopping. The exact amplitudes are
\begin{equation}
\varphi_n(t)
=(-i)^n e^{-iat}
\frac{(n+1)\mathrm J_{n+1}(2bt)}
{bt},
\end{equation}
and therefore
\begin{equation}
P_n(t)
=
\frac{(n+1)^2}{b^2t^2}
\mathrm J_{n+1}^2(2bt).
\end{equation}
Here $\mathrm J_n(x)$ denotes the ordinary Bessel function of the first kind. The wavepacket spreads ballistically and escapes to large Krylov depth.

Using the fixed-$n$ large-time asymptotic form of the Bessel function,
\begin{equation}
\mathrm J_{n+1}(2bt)
\sim
\sqrt{\frac{1}{\pi bt}}
\cos(2bt - \cdots),
\end{equation}
one finds $P_n(t)\sim t^{-3}$. Consequently,
\begin{equation}
\Pi_n=0,
\qquad
\overline{\chi_n}
=
\lim_{T\to\infty}
\frac{1}{T}
\int_0^T P_n^2(t)\,dt
=0 .
\end{equation}
Thus no nontrivial stationary occupation or residual fluctuation profile remains at fixed Krylov depth.

The accumulated quantities behave as
\begin{equation}
\mathcal{P}_n \to \frac{2}{\pi b},
\qquad
\mathcal Q_n \sim \frac{4}{\pi^2 b}\frac{\log n}{n},
\qquad
\mathcal J_n \sim \frac{16b}{\pi^2}\frac{\log n}{n}.
\end{equation}
Consequently, $\mathcal Q_n$ and $\mathcal J_n$ vanish at large depth, but only logarithmically slowly. Their partial sums grow as $(\log N)^2$. No fixed low-depth region captures a finite fraction of the accumulated return intensity or current-fluctuation activity in the infinite-chain limit. The constant-hopping model has ballistic Krylov growth and transient passage through every depth, but no normalizable stationary Krylov measure and no stationary memory core.

\subsubsection{Square-root growth: accelerating wavepacket}

We next consider a Krylov chain with vanishing diagonal coefficients and square-root hopping,
\begin{equation}
a_n=0,
\qquad
b_n=\lambda\sqrt n.
\end{equation}
The exact amplitudes are coherent-state wavefunctions~\cite{Balasubramanian:2022tpr},
\begin{equation}
\varphi_n(t)
=
e^{-\lambda^2t^2/2}
\frac{(-i\lambda t)^n}{\sqrt{n!}},
\end{equation}
giving
\begin{equation}
P_n(t)
=
e^{-\lambda^2t^2}
\frac{(\lambda^2t^2)^n}{n!}.
\end{equation}
For every fixed $n$, $P_n(t)\rightarrow 0$ as $t\to\infty$, and hence
\begin{equation}
\Pi_n=0,
\qquad
\overline{\chi_n}=0.
\end{equation}
The transient quantities are
\begin{equation}
\mathcal{P}_n \sim \frac{1}{2\lambda\sqrt n},
\qquad
\mathcal Q_n \sim \frac{1}{4\sqrt{\pi}\lambda n},
\qquad
\mathcal J_n \to \frac{\lambda}{\sqrt{\pi}}.
\end{equation}
The mean Krylov depth grows quadratically in time, so the wavepacket accelerates through Krylov space. However, this faster spreading still does not produce a stationary memory profile. At each fixed depth the averaged occupation and fluctuation vanish. The transient quantities describe how the escaping wavepacket visits and crosses each depth; they do not define a normalized stationary distribution or a memory core.

\subsubsection{Linear growth: maximal Krylov spreading}

As a third example we study a Krylov chain with Lanczos coefficients
\begin{equation}
a_n = 0,
\qquad
b_n = \alpha\sqrt{n(n+h-1)},
\qquad
h,\alpha>0 .
\end{equation}
This model possesses an exact $SU(1,1)$ structure and provides an analytically tractable realization of hyperbolic spreading in Krylov space~\cite{Balasubramanian:2022tpr}. It is often used as a model for rapid complexity growth~\cite{Parker:2018yvk}. This example is important because asymptotically linear Lanczos growth is often associated with rapid Krylov spreading in chaotic systems with continuous spectra~\cite{Sachdev:1993bu,Kitaev:2015,Kitaev:2017aw, Balasubramanian:2022tpr,Rabinovici:2022beu}.

For simplicity, set $h=1$. The amplitudes are~\cite{Balasubramanian:2022tpr}
\begin{equation}
\varphi_n(t)
=
(-i)^n
\frac{\tanh^n(\alpha t)}
{\cosh(\alpha t)},
\end{equation}
and hence
\begin{equation}
P_n(t)
=
\frac{\tanh^{2n}(\alpha t)}
{\cosh^2(\alpha t)}.
\end{equation}
At late times and fixed $n$, $P_n(t)\sim 4e^{-2\alpha t}$, so
\begin{equation}
\Pi_n=0,
\qquad
\overline{\chi_n}=0.
\end{equation}
The transient quantities are
\begin{equation}
\mathcal{P}_n = \frac{1}{\alpha(2n+1)},
\qquad
\mathcal Q_n \sim \frac{1}{8\alpha n^2},
\qquad
\mathcal J_n \to \frac{\alpha}{2}.
\end{equation}

This example makes the distinction especially clear. The occupation probability at every fixed depth vanishes exponentially in time, yet the integrated current-fluctuation activity through each bond remains finite. Strong Krylov-current activity during escape does not imply a stationary activity core; it is a property of propagation, not of a diagonal-ensemble memory profile. Even the canonical linear-growth geometry does not produce the kind of stationary low-depth core found in the anomalous many-body states of Sec.~\ref{sec:numerics}.

\subsection{Stationary memory versus Krylov growth}

The solvable escaping geometries establish the essential distinction. Strong
spreading, rapid complexity growth, and large transient current activity do
not imply stationary memory. In all three examples,
\begin{equation}
\Pi_n=0,\qquad
\overline{\chi_n}=0,\qquad
\Gamma_n=0
\end{equation}
at every fixed depth.

The anomalous many-body states studied in Sec.~\ref{sec:numerics} are
different because their Krylov spaces are finite-dimensional and possess
normalizable stationary measures. More importantly, their stationary
residual weights are not merely present; they are concentrated in a common
low-depth region identified by three complementary diagnostics.

This comparison clarifies the meaning of a memory core. A stationary Krylov
measure is necessary, but it is not sufficient. Generic thermalizing states
also possess stationary occupation profiles, yet their residual fluctuation,
Gibbs-deviation, and current-activity weights are weak or broadly
distributed. The memory core requires their simultaneous concentration
within a compact Krylov region.

The comparison also explains why Krylov complexity alone cannot identify
memory. Complexity measures how far the wavepacket spreads on average,
whereas the cumulative memory-core diagnostics determine how stationary
nonthermal information is organized after dephasing. Krylov growth and
stationary memory are therefore distinct aspects of many-body dynamics.


\section{Conclusions}
\label{sec:conclusions}

In this work we introduced Krylov-space memory cores as stationary,
depth-resolved structures characterizing anomalous initial-state dynamics
in nonintegrable quantum many-body systems that otherwise thermalize. The
central idea is to distinguish the stationary probability background from
the physical content carried by that background. The stationary occupation
profile $\Pi_n$ determines where late-time probability resides in Krylov
space, while three complementary diagnostics---the residual fluctuation
profile $\overline{\chi_n}$, the Gibbs-deviation profile $d_n$, and the
Krylov-current fluctuation activity $\Gamma_n$---determine whether that
stationary region retains nonthermal information and remains dynamically
active.

The main result is the discovery of a common stationary Krylov organization
shared by different anomalous mechanisms. In the weakly thermalizing mixed-
field Ising chain, confinement-sensitive Ising dynamics, and the scarred PXP
model, anomalous initial states develop a compact low-depth region in which
fluctuation weight, Gibbs mismatch, and current-fluctuation activity are
simultaneously concentrated. Generic reference states evolving under the
same Hamiltonians do not exhibit this combination: their residual signals
are weak, broadly distributed, or fail to co-localize. The repeated
appearance of this structure across distinct microscopic mechanisms is the
central finding of this work.

The memory core is not defined by the stationary occupation profile alone.
The occupation profile provides the probability background, but an occupied
Krylov region does not necessarily carry memory: it may remain close to the
Gibbs reference, exhibit negligible temporal fluctuations, or support little
probability exchange. The active memory-core depth is therefore determined
by the cumulative scales of the three residual diagnostics,
\begin{equation}
n_{\rm mc}^{(q)}
=
\max\{
n_\chi^{(q)},n_d^{(q)},n_\Gamma^{(q)}
\}.
\end{equation}
The separation between $n_\Pi^{(q)}$ and $n_{\rm mc}^{(q)}$, when present,
reveals a stationary halo: a broader region carrying probability but much
less of the residual memory content.

The main numerical comparisons use the $95\%$ cumulative depth, while the
finite-size appendix repeats the analysis at $99\%$ to expose the influence
of the far tails. At the largest accessible size of each sequence, the
active memory scale remains much smaller than both the cyclic dimension and
the corresponding stationary occupation depth at the stricter threshold.
The full finite-size sequences nevertheless display substantial tail
sensitivity, particularly for the confinement-sensitive bubble state and for
the PXP model at $L=20$--$24$. We therefore use the $99\%$ results as a
tail-sensitivity test rather than as evidence for a uniform finite-size
scaling law. The strong hierarchy recovered at the largest sizes shows that
the principal large-size core--halo conclusion is not an artifact of using
the $95\%$ threshold.

For the PXP model, the large-dimensional production profiles incorporate
the symmetry-enforced repeated gaps analytically through the chiral-pair
construction and therefore do not depend on a numerical
$\varepsilon_\Delta$. Orbit-compressed complete equal-gap calculations at
$L=18,20,22,24$ provide an independent validation: over the sampled stable
intervals of $\varepsilon_\Delta$, the full $\overline{\chi_n}$ and
$\Gamma_n$ profiles remain unchanged within the reported numerical accuracy,
all four associated $95\%$ and $99\%$ cumulative depths are unchanged, and
the complete-gap and chiral-pair-resolved profiles agree within the adopted
validation tolerance.

The numerical results show that the memory core is not a consequence of
nonintegrability, Krylov growth, or stationarity alone. The auxiliary
integrable Ising comparison demonstrates that compact Krylov structures can
also arise from conserved quantities and state selection, but they have a
different physical origin. The memory core identified here concerns special
initial states embedded inside otherwise thermalizing nonintegrable systems.

The primary $q=0.95$ finite-size results further indicate a separation
between the active memory scale and the full cyclic Krylov dimension. In the
scarred PXP model, the compact active region appears inside a full
numerically resolved Krylov space much larger than the approximate
forward-scattering scale. In the chaotic Ising model, the weakly thermalizing
state exhibits a similarly shallow active region inside an exponentially
growing Krylov sector. The primary data are compatible with growth no faster
than linear in $L$ and with a decreasing fraction of the cyclic space,
\begin{equation}
n_{\rm mc}^{(0.95)}=O(L),
\qquad
\frac{n_{\rm mc}^{(0.95)}}{\mathcal D_0}\rightarrow0 ,
\end{equation}
provided that the observed trends persist. We do not assign an asymptotic
scaling law to the strongly nonmonotonic $q=0.99$ sequences, which are used
to quantify sensitivity to the far tails.

The comparison with escaping infinite Krylov geometries further clarifies
the nature of the memory core. Ballistic spreading, accelerated spreading,
and asymptotically linear Lanczos growth can produce strong Krylov
propagation without a normalizable stationary Krylov measure. A memory core
is therefore not a growth phenomenon; it is a stationary organization of
residual information inside a finite Krylov space.

The broader implication is that anomalous
nonthermalization can possess a
common stationary Krylov-space organization. 
Different microscopic
mechanisms---weak thermalization, confinement, and quantum many-body
scarring---retain information through different physical processes, yet
they generate the same stationary pattern: a compact, dynamically active
memory core embedded within a broader stationary occupation background.
This core-halo structure provides a common Krylov-space framework 
for comparing how different anomalous mechanisms retain 
structured nonthermal information.

\section*{Data Availability} The numerical data and 
custom code supporting the findings of this 
study are available from the corresponding author 
upon reasonable request.

\section*{Author contributions}

M.A. and M.J.V. conceived the study and developed the theoretical
framework. M.J.V. performed the numerical calculations. Both authors
analyzed and interpreted the results and wrote and revised the manuscript.

\section*{Competing interests}

The authors declare no competing interests.




\section*{Acknowledgments}
We would like to thank Mohammad Reza Tanhayi
for useful discussions.
We are also grateful to Seyed Hamed Aboutalebi for 
insightful discussions on the numerical 
computations and for providing access to 
the computational facilities of the Condensed 
Matter National Laboratory at IPM, where 
the numerical calculations reported 
in this work were performed.
M.~A. acknowledges support from the
Iran National Science Foundation (INSF) 
under Project No.~4023620.
The authors also acknowledge the use of 
ChatGPT (OpenAI) as an editorial assistant 
in improving the clarity and presentation 
of the manuscript.

\appendix

\section{Derivation and interpretation of the Krylov-current fluctuation activity}
\label{app:gamma_derivation}

In this appendix, we derive the exact spectral representation of the bond-resolved Krylov-current fluctuation activity and clarify the distinct roles of energy-level degeneracies and energy-gap degeneracies. We use the active spectral notation introduced in Sec.~\ref{sec:stationary}: the set $\mathcal E_0$ contains the distinct energies selected by the initial state, $\omega_E$ denotes the corresponding active spectral weight, and $|\Phi_E\rangle$ is the normalized active vector associated with the energy $E$. Exact degeneracies of the Hamiltonian spectrum, including a possibly high-dimensional zero-energy eigenspace, are already incorporated into this construction through the corresponding spectral projectors.

The primary definition
\begin{equation}
\Gamma_n
=
\overline{J_n^2(t)}
\label{eq:Gamma-primary-app}
\end{equation}
is valid without any assumption concerning either level degeneracies or degeneracies of the active energy gaps. The exact answer is obtained by coherently grouping all transitions with the same Bohr frequency. A compact expression involving only matrix elements of the diagonal ensemble follows when the nondegenerate-active-gap condition in Eq.~\eqref{eq:nondegenerate-active-gaps} is satisfied.

\subsection{Bond current and its vanishing mean}

In the standard Lanczos gauge, the hopping coefficients $b_n$ are real and non-negative. The probability current across the bond connecting Krylov sites $n-1$ and $n$ is
\begin{equation}
J_n(t)
=
2b_n\,\operatorname{Im}
\!\left[
\varphi_n^*(t)\varphi_{n-1}(t)
\right],
\qquad
1\leq n\leq \mathcal D_0-1.
\label{eq:Jn-app}
\end{equation}
Equivalently,
\begin{equation}
J_n(t)
=
\langle\psi(t)|\widehat J_n|\psi(t)\rangle,
\label{eq:Jn-operator-expectation-app}
\end{equation}
where
\begin{equation}
\widehat J_n
=
-i b_n
\left(
|n\rangle\langle n-1|
-
|n-1\rangle\langle n|
\right)
\label{eq:Jn-operator-app}
\end{equation}
is Hermitian. With this convention, the Krylov probabilities obey the continuity equation
\begin{equation}
\partial_t P_n(t)
=
J_n(t)-J_{n+1}(t),
\qquad
J_0=J_{\mathcal D_0}=0.
\label{eq:continuity-app}
\end{equation}

For every finite Krylov chain considered in the many-body calculations, the long-time mean current vanishes on every bond, independently of any spectral nondegeneracy assumption. Since $P_n(t)$ is bounded,
\begin{equation}
\overline{\partial_tP_n}
=
\lim_{T\rightarrow\infty}
\frac{P_n(T)-P_n(0)}{T}
=
0.
\end{equation}
Time averaging Eq.~\eqref{eq:continuity-app} gives $\overline{J_n}=\overline{J_{n+1}}$. The boundary conditions then imply
\begin{equation}
\overline{J_n(t)}
=
0
\label{eq:mean-current-zero-app}
\end{equation}
for every bond. Consequently, $\Gamma_n$ is also the temporal variance of the signed Krylov current. The vanishing mean expresses the long-time balance of forward and backward probability transfer; it does not imply that the bond is dynamically inactive.

\subsection{Exact gap-resolved spectral representation}

Using the active spectral decomposition,
\begin{equation}
|\psi(t)\rangle
=
\sum_{E\in\mathcal E_0}
\sqrt{\omega_E}\,
e^{-iEt}
|\Phi_E\rangle,
\label{eq:psi-active-app}
\end{equation}
the bond current may be written as
\begin{equation}
J_n(t)
=
\sum_{E,F\in\mathcal E_0}
\sqrt{\omega_E\omega_F}\,
e^{i(E-F)t}
J_{EF}^{(n)},
\label{eq:Jn-spectral-app}
\end{equation}
where
\begin{equation}
J_{EF}^{(n)}
=
\langle\Phi_E|\widehat J_n|\Phi_F\rangle.
\label{eq:JEF-definition-app}
\end{equation}

The restriction of the Hamiltonian to the cyclic Krylov space is represented by a real symmetric Jacobi matrix. Its active energy eigenvectors can therefore be chosen real in the Krylov basis. In this gauge, $\widehat J_n$ is purely imaginary and antisymmetric, implying
\begin{equation}
J_{EE}^{(n)}
=
0.
\label{eq:J-diagonal-zero-app}
\end{equation}

It is useful to group all ordered transitions carrying the same active energy gap. For every gap $\Delta$, define
\begin{equation}
\mathcal A_n(\Delta)
=
\sum_{\substack{E,F\in\mathcal E_0\\E-F=\Delta}}
\sqrt{\omega_E\omega_F}\,
J_{EF}^{(n)}.
\label{eq:gap-amplitude-app}
\end{equation}
The current then has the Fourier decomposition
\begin{equation}
J_n(t)
=
\sum_\Delta
\mathcal A_n(\Delta)e^{i\Delta t}.
\label{eq:Jn-gap-decomposition-app}
\end{equation}
Hermiticity gives $\mathcal A_n(-\Delta)=\mathcal A_n(\Delta)^*$. Moreover, $\mathcal A_n(0)=0$ because $\mathcal E_0$ contains distinct active energies and the diagonal current matrix elements vanish according to Eq.~\eqref{eq:J-diagonal-zero-app}.

Squaring Eq.~\eqref{eq:Jn-gap-decomposition-app} and taking the infinite-time average imposes the frequency-matching condition $\Delta'=-\Delta$. We therefore obtain
\begin{equation}
\Gamma_n
=
\sum_\Delta
\left|
\mathcal A_n(\Delta)
\right|^2
\label{eq:Gamma-gap-resolved-app}
\end{equation}
without any assumption about active-gap degeneracies. Equation~\eqref{eq:Gamma-gap-resolved-app} is the general expression for the Krylov-current fluctuation activity.

When several ordered energy pairs share the same gap, their matrix elements must be added coherently at fixed $\Delta$ before the absolute square is taken. This ordering of the operations is essential:
\begin{equation}
\left|
\sum_{E-F=\Delta}
\sqrt{\omega_E\omega_F}\,
J_{EF}^{(n)}
\right|^2
\neq
\sum_{E-F=\Delta}
\omega_E\omega_F
\left|
J_{EF}^{(n)}
\right|^2
\end{equation}
in general.

Under the nondegenerate-active-gap condition \eqref{eq:nondegenerate-active-gaps}, every nonzero active gap is associated with a unique ordered pair. In that case, Eq.~\eqref{eq:Gamma-gap-resolved-app} reduces to
\begin{equation}
\Gamma_n
=
\sum_{E,F\in\mathcal E_0}
\omega_E\omega_F
\left|
J_{EF}^{(n)}
\right|^2,
\label{eq:Gamma-energy-basis-app}
\end{equation}
where the terms with $E=F$ vanish according to Eq.~\eqref{eq:J-diagonal-zero-app}. If the active-gap condition is violated, Eq.~\eqref{eq:Gamma-gap-resolved-app}, or an equivalent direct long-time average of Eq.~\eqref{eq:Gamma-primary-app}, must be used instead of Eq.~\eqref{eq:Gamma-energy-basis-app}.

\subsection{Exact level degeneracies and symmetry-enforced gap degeneracies}

Exact degeneracy of energy levels and degeneracy of energy gaps are logically distinct. The former is incorporated directly into the active spectral construction, whereas the latter produces coherent interference between different transition channels and must be retained in the evaluation of $\Gamma_n$.

Let $\mathsf P_E$ denote the full spectral projector onto the eigenspace of the Hamiltonian with energy $E$. The active weight and active vector are defined by
\begin{equation}
\omega_E
=
\langle\psi_0|\mathsf P_E|\psi_0\rangle,
\qquad
|\Phi_E\rangle
=
\frac{\mathsf P_E|\psi_0\rangle}
{\sqrt{\omega_E}},
\qquad
\omega_E>0.
\label{eq:active-projector-app}
\end{equation}
These definitions remain valid for arbitrary $\operatorname{rank}\mathsf P_E$. For a pure initial state, only the single direction $\mathsf P_E|\psi_0\rangle$ inside the degenerate eigenspace is dynamically active; all vectors orthogonal to it have zero overlap with the initial state and do not belong to the cyclic Krylov subspace. In exact arithmetic,
\begin{equation}
\mathcal D_0
=
|\mathcal E_0|,
\end{equation}
namely, the cyclic dimension equals the number of distinct active energies.

The exact time-averaged density matrix is correspondingly
\begin{equation}
\rho_{\rm diag}
=
\sum_{E\in\mathcal E_0}
\mathsf P_E
|\psi_0\rangle\langle\psi_0|
\mathsf P_E
=
\sum_{E\in\mathcal E_0}
\omega_E
|\Phi_E\rangle\langle\Phi_E|.
\label{eq:rho-diag-degenerate-app}
\end{equation}
Thus $\rho_{\rm diag}$ preserves the component of the initial-state coherence that lies within an exactly degenerate eigenspace. In particular, a high-dimensional zero-energy manifold contributes through the single normalized active vector $|\Phi_0\rangle$, provided that $\omega_0>0$.

Level degeneracies associated with commuting symmetries can be treated by first restricting the Hamiltonian to the symmetry sector selected by the initial state. This removes multiplicities arising solely from distinct commuting sectors. It does not, however, remove equalities among Bohr frequencies within the resolved active sector. Symmetry resolution and active-gap resolution are therefore separate steps.

To make this distinction explicit, define
\begin{equation}
g_{EF}^{(n)}
=
\sqrt{\omega_E\omega_F}\,
J_{EF}^{(n)}
\end{equation}
and the transition set
\begin{equation}
\mathscr{T}_\Delta
=
\left\{
(E,F)\in\mathcal E_0^2:
E-F=\Delta
\right\}.
\end{equation}
Equation~\eqref{eq:Gamma-gap-resolved-app} may then be decomposed as
\begin{equation}
\Gamma_n
=
\Gamma_n^{\rm inc}
+
\Gamma_n^{\rm coh},
\label{eq:Gamma-inc-coh-app}
\end{equation}
where
\begin{equation}
\Gamma_n^{\rm inc}
=
\sum_\Delta
\sum_{(E,F)\in\mathscr{T}_\Delta}
\left|
g_{EF}^{(n)}
\right|^2
=
\sum_{E,F\in\mathcal E_0}
\omega_E\omega_F
\left|
J_{EF}^{(n)}
\right|^2
\label{eq:Gamma-incoherent-app}
\end{equation}
is the incoherent transition contribution, and
\begin{equation}
\Gamma_n^{\rm coh}
=
2\,\operatorname{Re}
\sum_\Delta
\sum_{\substack{
p,q\in\mathscr T_\Delta\\
p<q
}}
g_p^{(n)}
g_q^{(n)*}
\label{eq:Gamma-gap-coherence-app}
\end{equation}
contains the interference between distinct ordered transitions with the same gap. Here $p<q$ denotes any fixed ordering used only to count each unordered pair once.

The quantity $\Gamma_n^{\rm coh}$ is a gap-coherence contribution. It is not encoded in the diagonal ensemble alone because it arises from correlations between off-diagonal transition amplitudes oscillating at the same nonzero frequency. The nondegenerate-active-gap condition is a sufficient condition for
\begin{equation}
\Gamma_n^{\rm coh}
=
0,
\end{equation}
although the coherent contribution may also vanish in special gap-degenerate cases through exact cancellations.

\subsubsection{Chiral symmetry in the PXP model}

The periodic PXP Hamiltonian provides an important example in which repeated gaps are enforced by symmetry even after all commuting spatial symmetries have been resolved. Let $\mathcal C$ denote the chiral operator satisfying
\begin{equation}
\mathcal C^2=\mathbb I,
\qquad
\mathcal C H\mathcal C=-H.
\label{eq:chiral-symmetry-app}
\end{equation}
For an initial state with definite chiral parity,
\begin{equation}
\mathcal C|\psi_0\rangle
=
c_0|\psi_0\rangle,
\qquad
c_0=\pm1,
\end{equation}
the spectral projectors obey $\mathsf P_{-E}=\mathcal C\mathsf P_E\mathcal C$. It follows that
\begin{equation}
\omega_{-E}
=
\omega_E,
\qquad
|\Phi_{-E}\rangle
=
c_0\mathcal C|\Phi_E\rangle.
\label{eq:chiral-active-pairing-app}
\end{equation}

The Krylov vectors alternate in chiral parity:
\begin{equation}
\mathcal C|n\rangle
=
c_0(-1)^n|n\rangle.
\label{eq:krylov-chiral-parity-app}
\end{equation}
Consequently, the diagonal Lanczos coefficients vanish, $a_n=0$, and the active-energy components satisfy
\begin{equation}
u_{-E,m}
=
(-1)^m u_{E,m},
\qquad
u_{Em}
=
\langle m|\Phi_E\rangle.
\label{eq:chiral-component-relation-app}
\end{equation}

The current operator is odd under the chiral transformation,
\begin{equation}
\mathcal C\widehat J_n\mathcal C
=
-\widehat J_n.
\end{equation}
Using Eq.~\eqref{eq:chiral-active-pairing-app} gives
\begin{align}
J_{-F,-E}^{(n)}
&=
-\,
J_{FE}^{(n)}
\nonumber\\
&=
J_{EF}^{(n)},
\label{eq:chiral-current-pairing-app}
\end{align}
where the second equality follows because, in the real Lanczos gauge, $J_{EF}^{(n)}$ is purely imaginary and $J_{FE}^{(n)}=-J_{EF}^{(n)}$.

Now consider the map
\begin{equation}
\mathscr S:
(E,F)\longmapsto(-F,-E).
\label{eq:chiral-gap-map-app}
\end{equation}
It preserves the energy gap:
\begin{equation}
(-F)-(-E)
=
E-F.
\end{equation}
Moreover, Eqs.~\eqref{eq:chiral-active-pairing-app} and \eqref{eq:chiral-current-pairing-app} imply
\begin{equation}
g_{-F,-E}^{(n)}
=
g_{EF}^{(n)}.
\label{eq:chiral-transition-amplitude-app}
\end{equation}
Therefore, the two transitions related by $\mathscr S$ contribute constructively to the same gap amplitude.

The transition set $\mathscr T_\Delta$ can be partitioned into fixed points of $\mathscr S$ and two-element chiral orbits. The fixed points satisfy $F=-E$. Let $\mathscr T_\Delta^{\rm fix}$ denote the fixed-point subset and let $\mathscr R_\Delta$ contain one representative from every two-element orbit. Then
\begin{equation}
\mathcal A_n(\Delta)
=
\sum_{(E,F)\in\mathscr T_\Delta^{\rm fix}}
g_{EF}^{(n)}
+
2
\sum_{(E,F)\in\mathscr R_\Delta}
g_{EF}^{(n)}
\label{eq:PXP-gap-amplitude-app}
\end{equation}
for a chiral-symmetric active sector with a definite-chiral-parity initial state. Any additional accidental gap degeneracies simply give further terms inside the same coherent sum.

For an isolated nonfixed chiral orbit, $\{(E,F),(-F,-E)\}$, the exact contribution at the corresponding gap is
\begin{equation}
\left|
g_{EF}^{(n)}
+
g_{-F,-E}^{(n)}
\right|^2
=
4\left|
g_{EF}^{(n)}
\right|^2.
\end{equation}
By contrast, the incoherent expression would assign
\begin{equation}
\left|
g_{EF}^{(n)}
\right|^2
+
\left|
g_{-F,-E}^{(n)}
\right|^2
=
2\left|
g_{EF}^{(n)}
\right|^2.
\end{equation}
Thus an isolated chiral pair doubles the contribution relative to the incoherent transition sum. When several chiral or accidental transition orbits share the same gap, interference among the different orbit amplitudes may be either constructive or destructive.

The zero-energy subspace is also covered by this construction. Although an exactly degenerate zero-energy manifold contributes only through the active vector $|\Phi_0\rangle$, the transitions $(E,0)$ and $(0,-E)$ have the same gap and form a nonfixed chiral pair. They must therefore be combined coherently according to Eq.~\eqref{eq:PXP-gap-amplitude-app}.

\subsubsection{Numerical implementation}

In the numerical PXP calculation, the Hamiltonian is first restricted to
the relevant commuting-symmetry blocks.  Distinct active energies are
identified with the level-resolution tolerance $\varepsilon_E$, and the
retained active spectrum and weights are explicitly symmetrized under
$E\leftrightarrow -E$.  The large-$\mathcal D_0$ values of
$\overline{\chi_n}$ and $\Gamma_n$ are then obtained from the
chiral-pair-resolved expressions, which sum the two members of every
symmetry-enforced transition orbit coherently.  This treatment also
includes the pairs $(E,0)$ and $(0,-E)$ associated with the active
zero-energy component.

For the validation sizes $L=18,20,22,24$, we additionally construct an
orbit-compressed complete set of nonzero active-energy differences and group
distinct chiral-orbit representatives with a numerical gap tolerance
$\varepsilon_\Delta$. Within each group the transition amplitudes are summed
before taking the absolute square, as required by the exact gap-resolved
expressions for $\Gamma_n$ and $\overline{\chi_n}$ derived above. The
complete-gap result is compared with the scalable chiral result over a range
of $\varepsilon_\Delta$ values. This comparison tests both the numerical
identification of equal gaps and the possible influence of additional
accidental gap degeneracies. The larger production results do not depend on
$\varepsilon_\Delta$; their validity is supported by the finite-size
agreement reported in Appendix~\ref{app:pxp_gap_validation}.

For completeness, the same distinction applies to the residual probability fluctuation profile. Defining
\begin{equation}
\mathcal B_n(\Delta)
=
\sum_{\substack{E,F\in\mathcal E_0\\E-F=\Delta}}
\sqrt{\omega_E\omega_F}\,
u_{En}u_{Fn},
\end{equation}
one has
\begin{equation}
\overline{\chi_n}
=
\sum_{\Delta\neq0}
\left|
\mathcal B_n(\Delta)
\right|^2.
\label{eq:chi-gap-resolved-app}
\end{equation}
Thus repeated active gaps must also be treated coherently in the numerical evaluation of $\overline{\chi_n}$.

\subsection{Reduction to diagonal-ensemble matrix elements}

The incoherent contribution $\Gamma_n^{\rm inc}$ in Eq.~\eqref{eq:Gamma-incoherent-app} can be reduced to matrix elements of the diagonal ensemble. Let
\begin{equation}
u_{Em}
=
\langle m|\Phi_E\rangle
\end{equation}
denote the real Krylov components of the active energy vectors. From Eq.~\eqref{eq:Jn-operator-app},
\begin{equation}
J_{EF}^{(n)}
=
-i b_n
\left(
u_{En}u_{F,n-1}
-
u_{E,n-1}u_{Fn}
\right).
\label{eq:JEF-components-app}
\end{equation}
Substitution into Eq.~\eqref{eq:Gamma-incoherent-app} yields
\begin{align}
\Gamma_n^{\rm inc}
={}&
b_n^2
\sum_{E,F\in\mathcal E_0}
\omega_E\omega_F
\left(
u_{En}u_{F,n-1}
-
u_{E,n-1}u_{Fn}
\right)^2
\nonumber\\
={}&
2b_n^2
\Bigg[
\left(
\sum_{E\in\mathcal E_0}
\omega_Eu_{En}^2
\right)
\left(
\sum_{F\in\mathcal E_0}
\omega_Fu_{F,n-1}^2
\right)
\nonumber\\
&\hspace{3.2cm}
-
\left(
\sum_{E\in\mathcal E_0}
\omega_Eu_{En}u_{E,n-1}
\right)^2
\Bigg].
\label{eq:Gamma-expanded-app}
\end{align}

The first two sums are the stationary occupations:
\begin{equation}
\sum_{E\in\mathcal E_0}
\omega_Eu_{Em}^2
=
\langle m|\rho_{\rm diag}|m\rangle
=
\Pi_m.
\end{equation}
In the real Lanczos gauge,
\begin{equation}
\sum_{E\in\mathcal E_0}
\omega_Eu_{En}u_{E,n-1}
=
\langle n|\rho_{\rm diag}|n-1\rangle.
\end{equation}
Restoring a phase-gauge-invariant form for the off-diagonal matrix element, we find
\begin{equation}
\Gamma_n^{\rm inc}
=
2b_n^2
\left[
\Pi_n\Pi_{n-1}
-
\left|
\langle n|\rho_{\rm diag}|n-1\rangle
\right|^2
\right].
\label{eq:Gamma-incoherent-diag-app}
\end{equation}

Under the nondegenerate-active-gap condition, $\Gamma_n^{\rm coh}=0$, so that $\Gamma_n=\Gamma_n^{\rm inc}$. Equation~\eqref{eq:Gamma-incoherent-diag-app} then becomes
\begin{equation}
\Gamma_n
=
2b_n^2
\left[
\Pi_n\Pi_{n-1}
-
\left|
\langle n|\rho_{\rm diag}|n-1\rangle
\right|^2
\right]
\label{eq:GammaExact-app}
\end{equation}
and is the compact expression quoted in the main text for sectors satisfying the nondegenerate-active-gap condition.

When active gaps are degenerate, the right-hand side of Eq.~\eqref{eq:GammaExact-app} gives only the incoherent part of the activity. The full result is instead
\begin{equation}
\Gamma_n
=
2b_n^2
\left[
\Pi_n\Pi_{n-1}
-
\left|
\langle n|\rho_{\rm diag}|n-1\rangle
\right|^2
\right]
+
\Gamma_n^{\rm coh},
\label{eq:Gamma-generalized-diag-app}
\end{equation}
where $\Gamma_n^{\rm coh}$ is given by Eq.~\eqref{eq:Gamma-gap-coherence-app}. This term contains information about degenerate transition frequencies that cannot be reconstructed from $\rho_{\rm diag}$ alone.

The incoherent contribution can also be written as
\begin{equation}
\Gamma_n^{\rm inc}
=
2b_n^2
\det
\begin{pmatrix}
\langle n-1|\rho_{\rm diag}|n-1\rangle
&
\langle n-1|\rho_{\rm diag}|n\rangle
\\
\langle n|\rho_{\rm diag}|n-1\rangle
&
\langle n|\rho_{\rm diag}|n\rangle
\end{pmatrix}.
\label{eq:Gamma-principal-minor-app}
\end{equation}
Thus, in the nondegenerate-active-gap case, the full activity is proportional to the determinant of the two-site principal minor of the diagonal ensemble associated with the bond $(n-1,n)$. In a gap-degenerate problem, this determinant describes the incoherent baseline, while equal-frequency transition interference supplies the additional term $\Gamma_n^{\rm coh}$.

Because $\rho_{\rm diag}$ is positive semidefinite, the Cauchy--Schwarz inequality gives
\begin{equation}
\left|
\langle n|\rho_{\rm diag}|n-1\rangle
\right|^2
\leq
\Pi_n\Pi_{n-1}.
\end{equation}
Consequently,
\begin{equation}
0
\leq
\Gamma_n^{\rm inc}
\leq
2b_n^2\Pi_n\Pi_{n-1}.
\label{eq:Gamma-incoherent-bound-app}
\end{equation}
When the nondegenerate-active-gap condition holds, this becomes the sharper bound on the full activity,
\begin{equation}
0
\leq
\Gamma_n
\leq
2b_n^2\Pi_n\Pi_{n-1}.
\label{eq:Gamma-bound-app}
\end{equation}

For arbitrary gap degeneracies, a general bound follows directly from Eq.~\eqref{eq:Jn-app}:
\begin{equation}
J_n^2(t)
\leq
4b_n^2P_n(t)P_{n-1}(t).
\end{equation}
Therefore,
\begin{equation}
0
\leq
\Gamma_n
\leq
4b_n^2
\overline{P_n(t)P_{n-1}(t)}
\leq
b_n^2
\left(
\Pi_n+\Pi_{n-1}
\right).
\label{eq:Gamma-general-bound-app}
\end{equation}
In the last step, we used $4xy\leq(x+y)^2\leq x+y$ for $x,y\geq0$ and $x+y\leq1$.

\subsection{Physical interpretation}

The neighboring occupations and the hopping $b_n$ determine the maximum current scale available on a Krylov bond, but they do not uniquely determine the realized activity. In the nondegenerate-gap case, stationary nearest-neighbor coherence reduces the activity from its occupation-controlled upper bound. In the gap-degenerate case, coherent interference between transition channels carrying the same Bohr frequency provides an additional contribution that may enhance or suppress the incoherent baseline.

The quantity $\Gamma_n$ should not be interpreted as physical-space transport, directed transport, or, by itself, as a thermalization measure. Because it is the variance of a signed current, it contains no information about the net direction of probability flow. A large value indicates persistent back-and-forth probability exchange across a Krylov bond, which may remain confined to a compact region of the Krylov chain. It is in this precise sense that $\Gamma_n$ measures long-time Krylov-current fluctuation activity.



\section{Auxiliary integrable Ising comparison}
\label{app:integrable_ising}

This appendix presents an auxiliary comparison with the integrable limit of
the Ising chain. These results are not used as evidence for the memory-core
phenomenon identified in the main text. Instead, they clarify the role of
integrability and demonstrate that compact Krylov-space structures are not
sufficient by themselves to characterize anomalous nonthermalization in
otherwise thermalizing systems.

The main examples of the paper concern special initial states evolving under
nonintegrable Hamiltonians where generic states thermalize, but selected
states retain nonthermal stationary structure. The integrable comparison has
a different physical origin: persistent structure is expected because the
dynamics is constrained by an extensive set of conserved quantities. The
purpose of this appendix is therefore not to classify integrability through
the Krylov diagnostics, but to show how similar stationary structures can
arise from different mechanisms.

We use the Ising Hamiltonian of Eq.~\eqref{eq:ising_hamiltonian_num} with
\begin{equation}
J=1,\qquad g=-1.05,\qquad h=0 .
\end{equation}
The initial states are the homogeneous product states
$|X+\rangle$, $|Y+\rangle$, and $|Z+\rangle$ defined in
Eq.~\eqref{eq:xyz_states_num}, together with a symmetry-matched random
product state. In this integrable setting these labels should not be
interpreted as thermalizing or nonthermalizing states; the random state is
used only as a generic reference within the same integrable dynamics.

\begin{figure}[h!]
\centering
\includegraphics[width=0.35\textwidth]{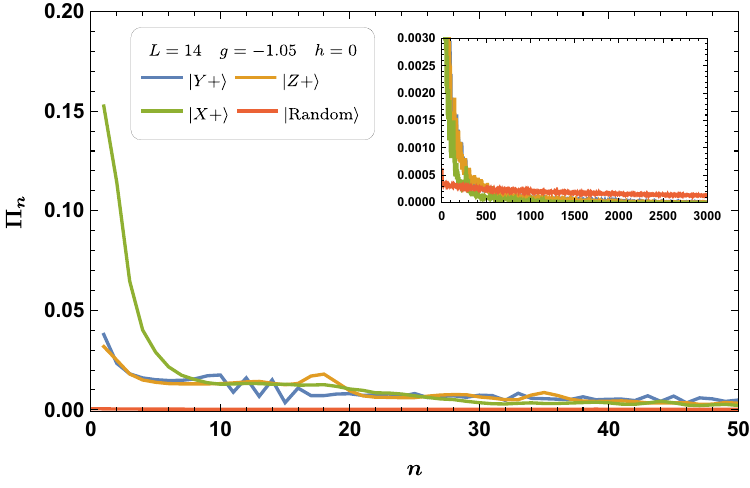}
\caption{
Stationary Krylov occupation profile $\Pi_n$ for the integrable Ising chain
($L=14$, $g=-1.05$, $h=0$). The homogeneous states exhibit strong
early-depth concentration, while the symmetry-matched random reference has a
broader stationary occupation profile. The figure illustrates that compact
stationary probability is not by itself a signature of anomalous
nonthermalization.
}
\label{fig:app_ising_pi}
\end{figure}

\paragraph{Stationary occupation profile.}
The stationary occupation profile $\Pi_n$ shown in Fig.~\ref{fig:app_ising_pi}
describes the late-time probability background in Krylov space. The
homogeneous states exhibit pronounced low-depth structure, while the random
state displays a broader distribution. This demonstrates that compact
stationary occupation is not unique to anomalous states in nonintegrable
systems. As in the main text, $\Pi_n$ specifies the stationary probability
background but does not by itself determine whether that probability carries
nonthermal residual structure.

\begin{figure}[h!]
\centering
\includegraphics[width=0.35\textwidth]{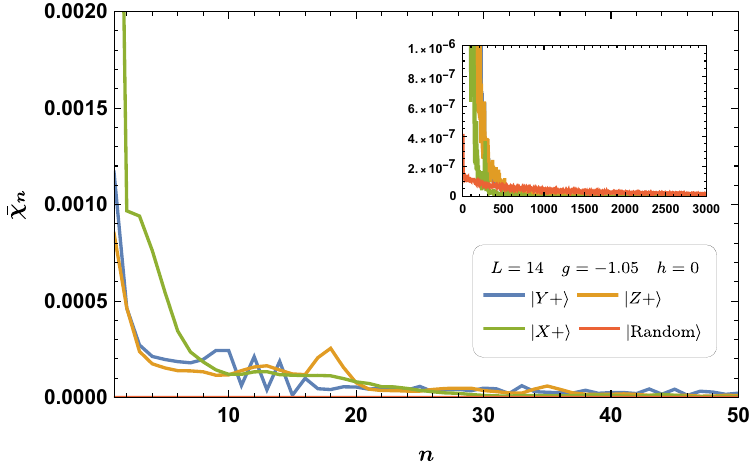}
\caption{
Residual fluctuation profile $\overline{\chi_n}$ for the integrable Ising
chain ($L=14$, $g=-1.05$, $h=0$). The homogeneous states show concentrated
early-depth fluctuations, while the random reference has weaker residual
fluctuation weight. The physical interpretation differs from the anomalous
nonintegrable cases because the structure originates from integrability.
}
\label{fig:app_ising_chi}
\end{figure}

\paragraph{Residual equilibration fluctuations.}
The residual fluctuation profile $\overline{\chi_n}$ is shown in
Fig.~\ref{fig:app_ising_chi}. The homogeneous states exhibit concentrated
early-depth fluctuation weight, whereas the random state has weaker
fluctuations. The origin of this structure differs from the nonintegrable
cases studied in the main text: here it follows from integrability and the
associated conserved quantities rather than from a special state embedded
inside a thermalizing spectrum.

\paragraph{Krylov-resolved Gibbs mismatch.}
In an integrable system the canonical Gibbs ensemble is generally not the
appropriate stationary reference because additional conserved quantities
constrain the dynamics. Therefore
\begin{equation}
d_n=\Pi_n-\Pi_n^{({\rm Gibbs})}
\end{equation}
should be interpreted here only as a mismatch relative to the canonical
Gibbs benchmark, not as a direct thermalization diagnostic. A generalized
Gibbs ensemble would provide the appropriate equilibrium comparison.

\begin{figure}[h!]
\centering
\includegraphics[width=0.35\textwidth]{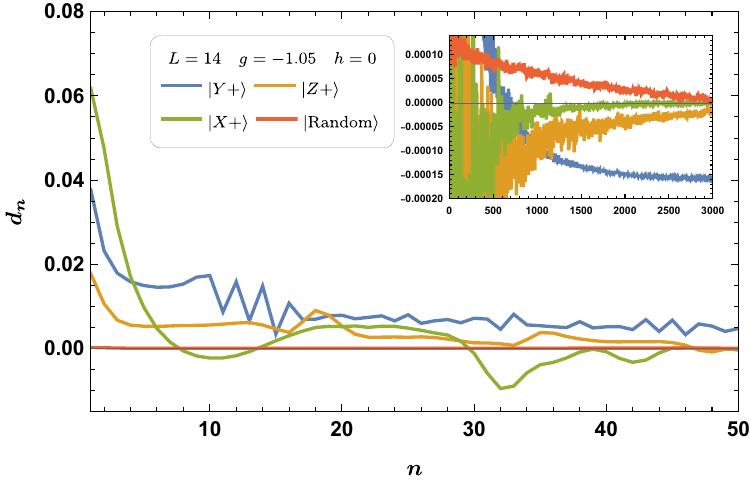}
\caption{
Signed Gibbs-mismatch profile
$d_n=\Pi_n-\Pi_n^{({\rm Gibbs})}$ for the integrable Ising chain
($L=14$, $g=-1.05$, $h=0$). The profile measures deviation from the
canonical Gibbs reference only. The homogeneous states exhibit stronger
early-depth mismatch, while the random reference remains closer to the
canonical benchmark.
}
\label{fig:app_ising_d}
\end{figure}

\begin{figure}[h!]
\centering
\includegraphics[width=0.35\textwidth]{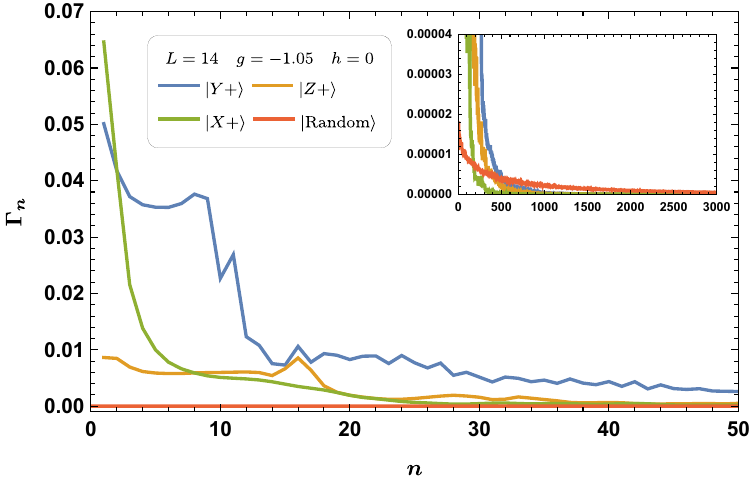}
\caption{
Krylov-current fluctuation activity $\Gamma_n$ for the integrable Ising
chain ($L=14$, $g=-1.05$, $h=0$). The homogeneous states display
early-depth current activity, whereas the random reference has weaker
activity throughout the Krylov chain.
}
\label{fig:app_ising_gamma}
\end{figure}

\paragraph{Krylov-current fluctuation activity.}
The current-fluctuation activity $\Gamma_n$ in
Fig.~\ref{fig:app_ising_gamma} measures persistent probability exchange
between neighboring Krylov depths. As in the main text, it is not a measure
of physical transport. The homogeneous states display stronger early-depth
activity, while the random reference has weaker activity.

\begin{figure}[h!]
\centering
\includegraphics[width=0.35\textwidth]{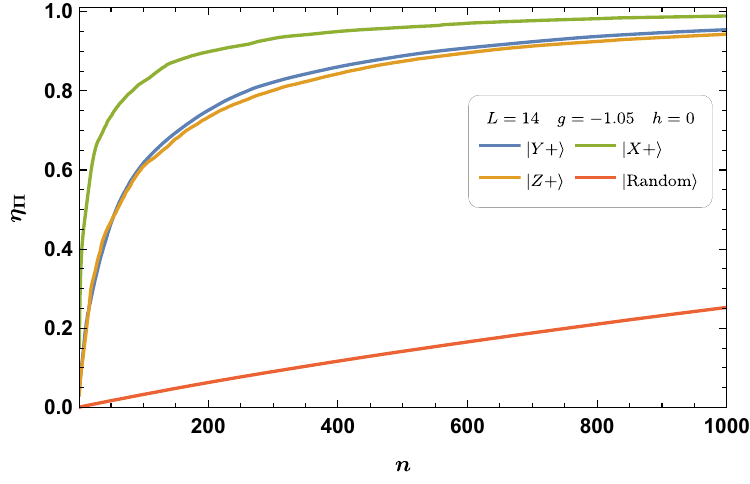}
\vspace{0.5em}
\includegraphics[width=0.35\textwidth]{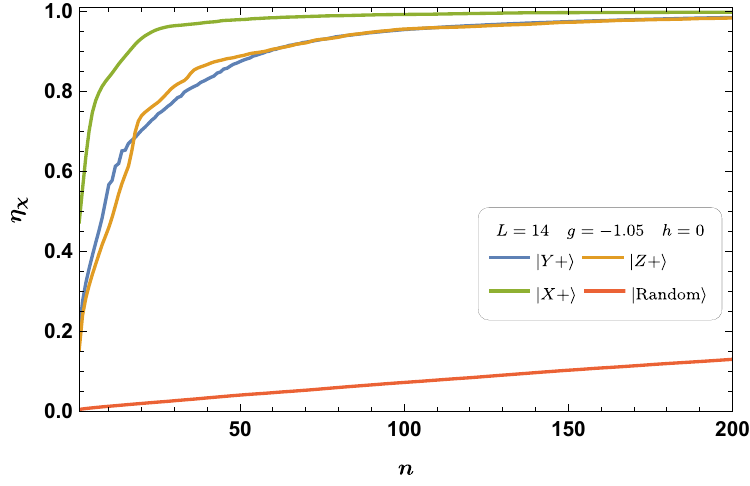}
\vspace{0.5em}
\includegraphics[width=0.35\textwidth]{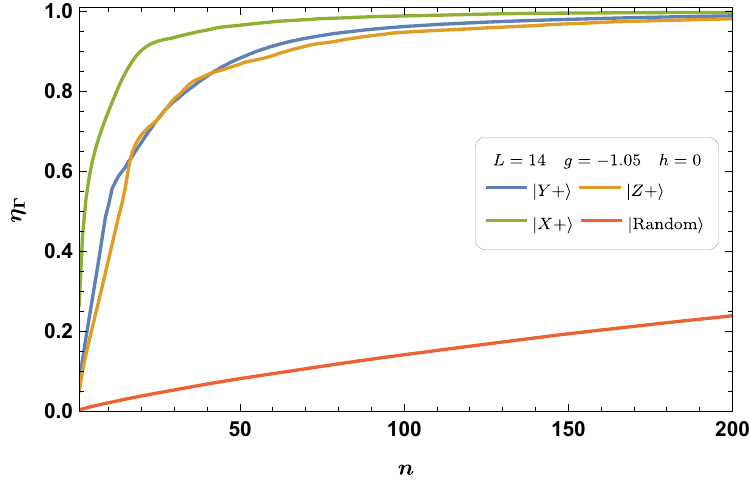}
\caption{Cumulative probability $\eta_\Pi$,
cumulative fluctuation profile $\eta_\chi$ and cumulative
current-fluctuation activity profile $\eta_\Gamma$ for the integrable
Ising chain ($L=14$, $g=-1.05$, $h=0$). The homogeneous states accumulate
their residual weights near the Krylov origin, whereas the random reference
shows a broader accumulation. These compact structures arise from
integrability and should not be confused with the anomalous memory cores
identified in otherwise thermalizing nonintegrable systems.
}
\label{fig:app_ising_cumulative}
\end{figure}

\paragraph{Cumulative diagnostics and comparison with memory cores.}
The cumulative fluctuation and current-activity profiles are shown in
Fig.~\ref{fig:app_ising_cumulative}. The homogeneous states exhibit rapid
saturation, indicating that their residual fluctuation and current activity
are concentrated near the Krylov origin. The random state accumulates these
weights more gradually.

We do not interpret these results as evidence for the memory-core
phenomenology of the main text. The compact residual structures observed
here arise in an integrable system where conservation laws constrain the
dynamics. Moreover, the ordinary Gibbs reference is not the appropriate
equilibrium description. The comparison instead demonstrates an important
distinction: similar Krylov-space geometries can emerge from different
physical origins.

The integrable comparison therefore serves as a control case. It shows that
compact Krylov-space structure is neither unique to nonintegrable anomalous
states nor sufficient by itself to define a memory core. The central result
of the main text is more specific: in otherwise thermalizing nonintegrable
systems, special initial states develop compact regions where residual
fluctuation, Gibbs mismatch, and current-fluctuation activity become
simultaneously concentrated, while generic states under the same Hamiltonian
do not.


\section{Symmetry bounds and active-spectral counting of the PXP Krylov dimension}
\label{app:pxp_dimension_count}

This appendix explains how the cyclic Krylov dimension reported for the periodic PXP dynamics is obtained. It is important to distinguish three quantities: the dimension of the full constrained Hilbert space, the dimension of the smallest symmetry-resolved invariant subspace containing the initial state, and the dimension of the cyclic subspace actually generated by that state. The first two provide state-independent or symmetry-based upper bounds. The last is a property of the pair $(H,|\psi_0\rangle)$ and is fixed by the distinct energy eigenspaces that carry nonzero initial-state spectral weight.

We first state the exact finite-dimensional result and then explain how it is resolved numerically for the periodic $L=28$ PXP N\'eel state.

\subsection{Exact active spectral dimension}

Let
\begin{equation}
H
=
\sum_{E\in\operatorname{spec}(H)}
E\,\mathsf P_E
\label{eq:app_pxp_spectral_resolution}
\end{equation}
be the spectral resolution of a finite-dimensional Hermitian Hamiltonian, where $\mathsf P_E$ projects onto the complete eigenspace associated with the distinct energy $E$. For a normalized initial state $|\psi_0\rangle$, define the active spectral set
\begin{equation}
\mathcal E_0
=
\left\{
E\in\operatorname{spec}(H):
\mathsf P_E|\psi_0\rangle\neq0
\right\},
\label{eq:app_pxp_active_set}
\end{equation}
together with
\begin{equation}
\omega_E
=
\langle\psi_0|\mathsf P_E|\psi_0\rangle,
\qquad
|\Phi_E\rangle
=
\frac{\mathsf P_E|\psi_0\rangle}{\sqrt{\omega_E}},
\qquad E\in\mathcal E_0 .
\label{eq:app_pxp_active_data}
\end{equation}
Repeated application of the Hamiltonian gives
\begin{equation}
H^m|\psi_0\rangle
=
\sum_{E\in\mathcal E_0}
E^m\sqrt{\omega_E}\,|\Phi_E\rangle .
\label{eq:app_pxp_Hm}
\end{equation}
Equation~\eqref{eq:app_pxp_Hm} shows that every Krylov vector belongs to $\operatorname{span}\{|\Phi_E\rangle:E\in\mathcal E_0\}$. Conversely, polynomial interpolation on the finite set of distinct active energies expresses every $|\Phi_E\rangle$ as a linear combination of Krylov vectors. Therefore
\begin{align}
\mathcal K
=&
\operatorname{span}
\left\{
|\Phi_E\rangle:E\in\mathcal E_0
\right\},
\cr
\mathcal D_0^{\rm exact}
=&
|\mathcal E_0|
=
\#\left\{
E:\mathsf P_E|\psi_0\rangle\neq0
\right\}.
\label{eq:app_pxp_exact_cyclic_dimension}
\end{align}
This is the exact cyclic spectral dimension of the pair $(H,|\psi_0\rangle)$~\cite{Lanczos1950,Saad2003,LiesenStrakos2013}.

An exactly degenerate eigenspace contributes at most one Krylov direction: the direction selected by $\mathsf P_E|\psi_0\rangle$. If this projection vanishes, the entire eigenspace is inactive. A degenerate zero-energy eigenspace is treated in precisely the same way and does not require a separate definition.

The same active data define the discrete spectral measure
\begin{equation}
\mu_{\psi_0}(\lambda)
=
\sum_{E\in\mathcal E_0}
\omega_E\,\delta(\lambda-E).
\label{eq:app_pxp_spectral_measure}
\end{equation}
The Lanczos coefficients are the Jacobi recurrence coefficients associated with this measure~\cite{GolubMeurant2010,Meurant2006,MeurantStrakos2006}. Consequently, the active spectral measure determines both the cyclic dimension and the tridiagonal Krylov Hamiltonian. This is more reliable than identifying $\mathcal D_0$ from the stopping index of an unreorthogonalized finite-precision Lanczos recursion.

\subsection{Symmetry-sector bounds}

For a periodic PXP chain, the constrained Hilbert-space dimension is
\begin{equation}
\dim\mathcal H_{\rm PXP}^{\rm per}(L)
=
F_{L-1}+F_{L+1},
\label{eq:app_pxp_full_dimension}
\end{equation}
where $F_m$ denotes the Fibonacci sequence. Thus
\begin{equation}
\dim\mathcal H_{\rm PXP}^{\rm per}(28)
=
F_{27}+F_{29}
=
710647 .
\label{eq:app_pxp_L28_full_dimension}
\end{equation}

Let $T$ denote translation by one site. Since
\begin{equation}
T^2|Z_2\rangle=|Z_2\rangle,
\end{equation}
the cyclic dynamics is contained in
\begin{equation}
\mathcal H_{T^2=1}
=
\mathcal H_{k=0}\oplus\mathcal H_{k=\pi}.
\label{eq:app_pxp_T2_sector}
\end{equation}
The standard character projection gives
\begin{equation}
\dim\mathcal H_{k=0}=25415,
\qquad
\dim\mathcal H_{k=\pi}=25410,
\end{equation}
and hence
\begin{equation}
\dim\mathcal H_{T^2=1}=50825.
\label{eq:app_pxp_T2_dimension}
\end{equation}
This number is a symmetry-sector upper bound, not the Krylov dimension.

The two momentum components of $|Z_2\rangle$ also have definite parity under a reflection $R$ satisfying $RTR=T^{-1}$. The reflection blocks selected by the N\'eel components have dimensions
\begin{equation}
\dim\mathcal H_{k=0,R_{Z_2}}=13201,
\qquad
\dim\mathcal H_{k=\pi,R_{Z_2}}=13010 .
\label{eq:app_pxp_reflection_blocks}
\end{equation}
The smallest symmetry-resolved invariant support containing the N\'eel-state dynamics therefore has dimension
\begin{equation}
13201+13010=26211 .
\label{eq:app_pxp_reflection_support}
\end{equation}
This is a sharper upper bound, but it still contains multiple vectors inside degenerate energy eigenspaces and may contain entire eigenspaces with vanishing N\'eel-state overlap.

\subsection{Zero-mode correction}

The PXP Hamiltonian possesses the chiral symmetry
\begin{equation}
\mathcal C=(-1)^{N_{\rm exc}},
\qquad
\{\mathcal C,H_{\rm PXP}\}=0,
\label{eq:app_pxp_chiral_symmetry}
\end{equation}
which pairs nonzero energies as $E\leftrightarrow-E$ and permits exact zero-energy states. For a bipartite Hamiltonian, the imbalance between the two chiral sublattices gives a lower bound on the zero-mode multiplicity \cite{Sutherland1986,Inui1994}; the corresponding PXP counting is discussed in Ref.~\cite{Buijsman:2022bkj}.

In the two reflection-resolved blocks of Eq.~\eqref{eq:app_pxp_reflection_blocks}, the chiral sublattice dimensions are
\begin{align}
(6667,6534)
&\qquad\text{for }(k=0,R_{Z_2}),
\nonumber\\
(6476,6534)
&\qquad\text{for }(k=\pi,R_{Z_2}).
\label{eq:app_pxp_chiral_sublattices}
\end{align}
The corresponding imbalances imply at least $133$ and $58$ zero modes, respectively, and the numerically resolved block spectra saturate these bounds, so the actual zero-mode counts are $N_0=133$ for the $k=0$ block and $N_0=58$ for the $k=\pi$ block.

The N\'eel state has nonzero total zero-energy weight through its $k=0$ component, whereas its selected $k=\pi$ component has no resolved zero-mode weight. Since all zero modes belong to the single energy $E=0$, their active projection contributes one Krylov direction, not one direction per zero mode. If zero-energy degeneracy were the only spectral reduction beyond the symmetry support, one would obtain
\begin{equation}
26211-(133+58)+1=26021 .
\label{eq:app_pxp_zero_mode_upper_bound}
\end{equation}
Equation~\eqref{eq:app_pxp_zero_mode_upper_bound} is therefore a zero-mode-corrected upper bound. It assumes the numerically resolved zero-mode multiplicities saturate the symmetry lower bounds, as verified for the two blocks, and that the $k=0$ zero-energy projection of the initial state is nonzero while the $k=\pi$ projection vanishes. It is not, by itself, the final state-dependent cyclic dimension.

\subsection{Numerical active-set construction}

The exact definition in Eq.~\eqref{eq:app_pxp_exact_cyclic_dimension} requires exact knowledge of both energy degeneracies and the support of the initial-state spectral measure. In floating-point arithmetic, however, one must determine numerically which energy levels are degenerate and which spectral weights are effectively zero. We therefore introduce explicit tolerances for resolving the active spectral measure. These tolerances are logically distinct from the gap-resolution tolerance introduced below: $\varepsilon_E$ and $\varepsilon_\omega$ determine the numerical active set and hence $\mathcal D_0$, whereas $\varepsilon_\Delta$ determines how repeated nonzero Bohr frequencies are grouped after the active set has already been fixed.

After diagonalizing the selected momentum--reflection blocks, their spectra are combined and numerically degenerate energies are merged using
\begin{equation}
\varepsilon_E=10^{-11}.
\label{eq:app_pxp_energy_tolerance}
\end{equation}
For each resulting energy group $g$, the total N\'eel-state spectral weight is
\begin{equation}
\omega_g
=
\sum_{\alpha\in g}
\left|
\langle E_\alpha|Z_2\rangle
\right|^2 .
\label{eq:app_pxp_group_weight}
\end{equation}
When $\omega_g>0$, the representative energy of the group is chosen as the spectral-weighted mean
\begin{equation}
\bar E_g
=
\frac{1}{\omega_g}
\sum_{\alpha\in g}
E_\alpha
\left|
\langle E_\alpha|Z_2\rangle
\right|^2 .
\label{eq:app_pxp_group_energy}
\end{equation}
For an exactly degenerate group, $\bar E_g$ equals the common exact energy; for a numerically merged group it is only a numerical representative. A group is retained as numerically active when
\begin{equation}
\omega_g>\varepsilon_\omega,
\qquad
\varepsilon_\omega=10^{-14}.
\label{eq:app_pxp_weight_tolerance}
\end{equation}
The retained weights are subsequently normalized to unit total weight before the Krylov recurrence is constructed. These tolerance values refer to the Hamiltonian normalization used throughout the main text. The resulting numerical active set and its dimension are
\begin{equation}
\mathcal E_0^{(\varepsilon_E,\varepsilon_\omega)}
=
\left\{
\bar E_g:\omega_g>\varepsilon_\omega
\right\},
\qquad
\mathcal D_0^{(\varepsilon_E,\varepsilon_\omega)}
=
\left|
\mathcal E_0^{(\varepsilon_E,\varepsilon_\omega)}
\right|.
\label{eq:app_pxp_numerical_active_set}
\end{equation}

For the $L=28$ N\'eel state, six additional numerical energy groups have total spectral weights below the cutoff. The discarded weights have
\begin{equation}
\max_{\text{discarded}}\omega_g \approx 7.9\times10^{-15},
\qquad
\min_{\text{retained}}\omega_g \approx 1.4\times10^{-14},
\end{equation}
Thus $\varepsilon_\omega=10^{-14}$ lies between the largest discarded and smallest retained group weights for this calculation. The numerically resolved active cyclic dimension at $L=28$ is therefore
\begin{equation}
\mathcal D_0^{(10^{-11},10^{-14})}
(|Z_2\rangle;L=28)
=
26015 .
\label{eq:app_pxp_L28_numerical_D0}
\end{equation}

The complete dimensional-reduction sequence is
\begin{equation}
710647
\;\longrightarrow\;
50825
\;\longrightarrow\;
26211
\;\longrightarrow\;
26021
\;\longrightarrow\;
26015 .
\label{eq:app_pxp_reduction_sequence}
\end{equation}
The first three numbers arise from the constrained Hilbert space and the successive exact spatial-symmetry reductions described above. The value $26021$ is the zero-mode-corrected spectral upper bound. The final step is state dependent and removes the numerical energy groups whose total N\'eel-state weights lie below $\varepsilon_\omega$.

The implementation records the retained and discarded spectral weights, the smallest candidate group weights, and the thresholds used to classify the numerical energy groups as active. The active-set construction was also tested under reasonable variations of $\varepsilon_E$ and $\varepsilon_\omega$. These variations change the numerically resolved value of $\mathcal D_0$ and the associated cumulative depths by only a few units, while leaving the low-depth profiles and the resulting core--halo separation unchanged within numerical accuracy. We therefore refer to $26015$ as the numerically resolved active cyclic dimension at the stated tolerances, rather than as an analytically proven exact integer. The exact mathematical cyclic dimension remains the quantity defined in Eq.~\eqref{eq:app_pxp_exact_cyclic_dimension}.
\subsection{Gap-resolution validation for the PXP diagnostics}
\label{app:pxp_gap_validation}

Once the numerical active set has been fixed, the evaluation of $\overline{\chi_n}$ and $\Gamma_n$ requires a separate identification of repeated nonzero Bohr frequencies. This operation does not alter $\mathcal E_0^{(\varepsilon_E,\varepsilon_\omega)}$ or $\mathcal D_0^{(\varepsilon_E,\varepsilon_\omega)}$.

The PXP chiral symmetry enforces the pairing
\begin{equation}
(E,F)\longleftrightarrow(-F,-E),
\label{eq:app_pxp_chiral_gap_orbit}
\end{equation}
whose two ordered transitions carry the same Bohr frequency. For a non-self-paired chiral orbit $\mathcal O=\{(E,F),(-F,-E)\}$, the chiral symmetry implies $g_{-F,-E}=g_{EF}$, so the orbit contributes
\begin{equation}
G_{\mathcal O}=g_{EF}+g_{-F,-E}=2g_{EF}
\end{equation}
to the gap amplitude. This symmetry-enforced contribution is incorporated analytically in the chiral-pair-resolved expressions used for the large-$\mathcal D_0$ production calculations.

To validate that additional numerical equal-gap classes do not modify the
resulting profiles, we perform orbit-compressed complete-gap calculations for
the PXP N\'eel state at $L=18,20,22,24$. These sizes extend the validation to
$\mathcal D_0=4569$ while remaining accessible to an independent complete-gap
treatment. Each chiral orbit of ordered transitions is represented once by
its amplitude $G_{\mathcal O}$. Distinct orbit representatives whose
numerical gaps coincide within $\varepsilon_\Delta$ are then combined
coherently before taking the absolute square. The procedure is therefore
complete with respect to all numerical equal-gap classes resolved at the
specified tolerance, while avoiding redundant enumeration of the
symmetry-related partners.

The effective gap tolerance is defined as
\begin{equation}
\varepsilon_\Delta
=
f_\Delta
\left[
\varepsilon_{\Delta,\mathrm{abs}}
+
\varepsilon_{\Delta,\mathrm{rel}}
\max\left(1,E_{\max}\right)
\right],
\label{eq:app_effective_gap_tolerance}
\end{equation}
with $E_{\max}
=
\max_{E\in\mathcal E_0}|E|$, and
\begin{equation}
\varepsilon_{\Delta,\mathrm{abs}}=10^{-10},
\qquad
\varepsilon_{\Delta,\mathrm{rel}}=10^{-12}.
\label{eq:app_gap_tolerance_components}
\end{equation}
The selected tolerance corresponds to $f_\Delta=1$. Stability is tested using
\begin{equation}
f_\Delta\in\{0.1,\,0.3,\,1,\,3,\,10\}.
\label{eq:app_gap_tolerance_factors}
\end{equation}

The definition of $\varepsilon_\Delta$ in Eq.~\eqref{eq:app_effective_gap_tolerance} addresses two distinct sources of numerical uncertainty. The absolute term $\varepsilon_{\Delta,\mathrm{abs}}$ sets a floor on the resolvable gap difference, preventing spurious gap splittings caused by rounding errors in the spectral data at small gap scales. The relative term $\varepsilon_{\Delta,\mathrm{rel}}\max(1,E_{\max})$ allows the clustering tolerance to scale with the characteristic magnitude of the spectral data, reflecting the growth of floating-point absolute errors with the magnitude of represented quantities. The factor $\max(1,E_{\max})$ prevents the tolerance from becoming artificially small when the spectrum lies entirely below unity. The scale factor $f_\Delta$ then probes the sensitivity of the observables to variations in the clustering resolution over a range spanning two orders of magnitude.

The positive- and negative-frequency sectors are clustered separately, so that a small nonzero gap is never merged with its opposite-frequency partner across $\Delta=0$. The $\Delta=0$ class is treated separately from both positive- and negative-frequency sectors. Within each sign sector, the sorted gaps are grouped using a maximum-diameter criterion: every member of a numerical gap class must lie within $\varepsilon_\Delta$ of the first member of that class. This prescription avoids both sign mixing and the transitive chaining that can arise in a nearest-neighbor or single-linkage construction.

To assess the sensitivity of the observables to the choice of $\varepsilon_\Delta$, we compute the relative $L^2$ difference
\begin{equation}
\delta_2[Q;f_\Delta]
=
\frac{
\left\|
Q(f_\Delta)-Q(1)
\right\|_2
}{
\left\|
Q(1)
\right\|_2
},
\qquad
Q\in\{\overline{\chi},\Gamma\},
\label{eq:app_gap_profile_distance}
\end{equation}
where the $L^2$ norm is taken over the full profile range
$n=0,\ldots,\mathcal D_0-1$. We also monitor the cumulative depths
$n_{\chi}^{(0.95)}$, $n_{\Gamma}^{(0.95)}$,
$n_{\chi}^{(0.99)}$, and $n_{\Gamma}^{(0.99)}$. For each sampled value of
$f_\Delta$, the maximum depth displacement relative to the selected-tolerance
result is
\[
\Delta n_{\max}(f_\Delta)
=
\max_{\substack{\ell\in\{\chi,\Gamma\}\\q\in\{0.95,0.99\}}}
\left|
 n_\ell^{(q)}(f_\Delta)-n_\ell^{(q)}(1)
\right|.
\]
A sampled tolerance is classified as stable when both profile changes are
smaller than $10^{-6}$, the complete-gap and chiral-pair-resolved profiles
agree to relative $L^2$ accuracy better than $10^{-8}$, and
$\Delta n_{\max}\leq3$.

\begin{table*}[t]
\centering
\small
\setlength{\tabcolsep}{4.0pt}
\resizebox{\textwidth}{!}{
\begin{tabular}{c c c c c c c c}
\hline\hline
Validation state
& $L$
& $\mathcal D_0$
& selected $\varepsilon_\Delta$
& sampled stable interval
& $\max_{f_\Delta}\delta_2(\overline\chi)$
& $\max_{f_\Delta}\delta_2(\Gamma)$
& maximum depth shift
\\
\hline
$|Z_2\rangle$ & 18 & 375 & $1.109\times10^{-10}$ & $[1.109\times10^{-11},\,1.109\times10^{-9}]$ & $0$ & $0$ & 0 \\
$|Z_2\rangle$ & 20 & 853 & $1.121\times10^{-10}$ & $[1.121\times10^{-11},\,1.121\times10^{-9}]$ & $1.175\times10^{-16}$ & $2.725\times10^{-17}$ & 0 \\
$|Z_2\rangle$ & 22 & 1937 & $1.133\times10^{-10}$ & $[1.133\times10^{-11},\,1.133\times10^{-9}]$ & $1.993\times10^{-12}$ & $4.938\times10^{-13}$ & 0 \\
$|Z_2\rangle$ & 24 & 4569 & $1.145\times10^{-10}$ & $[1.145\times10^{-11},\,1.145\times10^{-9}]$ & $8.359\times10^{-11}$ & $4.349\times10^{-11}$ & 0 \\
\hline\hline
\end{tabular}}
\caption{
Stability of the orbit-compressed complete equal-gap PXP calculation under
variation of $\varepsilon_\Delta$. The selected tolerance corresponds to
$f_\Delta=1$, and the maxima are taken over
$f_\Delta\in\{0.1,0.3,1,3,10\}$. Here $\delta_2$ is the relative $L^2$
difference from the selected-tolerance profile, while ``maximum depth
shift'' is the largest absolute change among
$n_{\chi}^{(0.95)}$, $n_{\Gamma}^{(0.95)}$,
$n_{\chi}^{(0.99)}$, and $n_{\Gamma}^{(0.99)}$.
The displayed intervals are the ranges covered by the sampled tolerance
factors and should not be interpreted as determinations of the outer plateau
boundaries. At every listed size, the complete-gap and chiral-pair-resolved
profiles agree within the validation tolerance $10^{-8}$.
}
\label{tab:app_pxp_gap_stability}
\end{table*}

All sampled values of $\varepsilon_\Delta$ satisfy the stability criteria at
every validation size in Table~\ref{tab:app_pxp_gap_stability}. The largest
relative profile variations occur at $L=24$ and are
$8.359\times10^{-11}$ for $\overline{\chi_n}$ and
$4.349\times10^{-11}$ for $\Gamma_n$, more than four orders of magnitude
below the adopted profile-stability threshold. None of the four cumulative
depths changes anywhere in the sampled intervals. At the selected
tolerances, the largest complete-gap versus chiral-pair relative differences
are $4.213\times10^{-12}$ for $\overline{\chi_n}$ and
$1.425\times10^{-12}$ for $\Gamma_n$. Thus additional numerically resolved
equal-gap classes produce no meaningful change in either the profile shapes
or their cumulative extents over the tested sizes and tolerance ranges. These
results validate the scalable chiral-pair-resolved expressions used for the
larger production calculations. Importantly, this validation concerns only
the treatment of repeated gaps; it does not modify the numerical active
dimension, which is determined exclusively by $\varepsilon_E$ and
$\varepsilon_\omega$.

\section{Finite-size behavior and threshold robustness of memory-core depths}
\label{app:core_scaling}

This appendix collects the finite-size data used to assess both the spatial behavior of the Krylov-space memory core and its sensitivity to the far tails of the cumulative profiles. Here $n$ denotes depth in the Krylov/Lanczos basis, whereas $\mathcal D_0$ denotes the total dimension of the cyclic Krylov subspace; these are distinct quantities. The former measures how far into the Krylov chain a specified fraction of cumulative weight extends, while the latter counts the total number of active Krylov directions.

For a threshold $q$, we define
\begin{equation}
 n_{\rm mc}^{(q)}
 =
 \max\left\{
 n_\chi^{(q)},
 n_d^{(q)},
 n_\Gamma^{(q)}
 \right\},
 \label{eq:app_nmc_definition}
\end{equation}
provided that all three integrated weights are numerically resolved.
Whether these weights are physically appreciable is assessed through
comparison with the matched reference states in the main text. No diagnostic
is selectively removed from the maximum in
Tables~\ref{tab:app_ising_core_scaling}--\ref{tab:app_pxp_core_scaling}. The
maximum is chosen so that the reported memory-core depth encloses the spatial
support of all three memory-related diagnostics rather than the depth of any
single diagnostic. Thus $n_{\rm mc}^{(q)}$ should be understood as an
operational enclosing depth for the three diagnostics; by itself, it is not a
diagnostic of anomalous memory. The occupation depth
\begin{equation}
 n_\Pi^{(q)}
 =
 \min\{n_c:\eta_\Pi(n_c)\geq q\}
 \label{eq:app_npi_definition}
\end{equation}
is reported separately because it characterizes the spatial extent of the stationary occupation profile rather than the active memory region.

We report both $q=0.95$ and $q=0.99$. The $95\%$ threshold is the primary value used in the main text, while the $99\%$ threshold probes the additional spatial extent required to increase the captured cumulative weight from $95\%$ to $99\%$. For any profile $\ell\in\{\Pi,\chi,d,\Gamma\}$, we define the tail-induced increase
\begin{equation}
 \Delta n_\ell^{(99-95)}
 =
 n_\ell^{(0.99)}-n_\ell^{(0.95)}\geq0.
 \label{eq:app_tail_increment}
\end{equation}
For the derived memory-core depth, the corresponding tail-induced increase is
\begin{equation}
 \Delta n_{\rm mc}^{(99-95)}
 =
 n_{\rm mc}^{(0.99)}-n_{\rm mc}^{(0.95)}\geq0.
\end{equation}
A compact $95\%$ depth accompanied by a much larger $99\%$ depth would indicate strong sensitivity to the far tails. Conversely, persistence of a strong hierarchy
\begin{equation}
 n_{\rm mc}^{(q)}\ll n_\Pi^{(q)},
 \qquad
 n_{\rm mc}^{(q)}\ll\mathcal D_0,
 \qquad q\in\{0.95,0.99\},
 \label{eq:app_two_threshold_hierarchy}
\end{equation}
quantified below by the corresponding ratios, would show that the core--halo separation is not an artifact of discarding the final $5\%$ of the cumulative profiles.

The finite-size tables below focus on spatial extent and therefore do not repeat the integrated weights $\mathcal X$, $\mathcal N$, and $\mathcal G$. These weights remain essential to the operational memory-core criterion and are therefore not omitted from the analysis; their values are reported in the main state-comparison tables and retained in the machine-readable numerical data.

\subsection{Weakly thermalizing Ising state}

We first consider $|X+\rangle$ in the chaotic mixed-field Ising chain,
\begin{equation}
 J=1,
 \qquad
 g=-1.05,
 \qquad
 h=0.5.
\end{equation}
The full Hilbert-space dimension is $2^L$. The reflection-even sector has dimension
\begin{equation}
 \dim\mathcal H_{R=+1}^{\rm Ising}(L)
 =
 \frac{1}{2}\left(2^L+2^{\lceil L/2\rceil}\right),
\end{equation}
but the cyclic dimension $\mathcal D_0$ is determined by the active spectral measure and need not equal the full sector dimension.

\begin{table*}[t]
\centering
\tiny
\setlength{\tabcolsep}{4.0pt}
\resizebox{\textwidth}{!}{
\begin{tabular}{c c c | c c c c c | c c c c c}
\hline\hline
&&&
\multicolumn{5}{c|}{$q=0.95$}
&
\multicolumn{5}{c}{$q=0.99$}
\\
$L$
& $\dim\mathcal H^{\rm Ising}$
& $\mathcal D_0$
& $n_\Pi^{(q)}$
& $n_\chi^{(q)}$
& $n_d^{(q)}$
& $n_\Gamma^{(q)}$
& $n_{\rm mc}^{(q)}$
& $n_\Pi^{(q)}$
& $n_\chi^{(q)}$
& $n_d^{(q)}$
& $n_\Gamma^{(q)}$
& $n_{\rm mc}^{(q)}$
\\
\hline
10 & 1024  & 528  & 205  & 27 & 15 & 30 & 30 & 384 & 98 & 25 & 88 & 98 \\
11 & 2048  & 1056 & 365  & 19 & 20 & 24 & 24 & 704 & 96 & 64 & 97 & 97 \\
12 & 4096  & 2080 & 657  & 20 & 4  & 24 & 24 & 1383 & 114 & 49 & 121 & 121 \\
13 & 8192  & 4160 & 1194 & 26 & 4  & 36 & 36 & 2653 & 145 & 67 & 160 & 160 \\
14 & 16384 & 8252 & 2053 & 25 & 3  & 32 & 32 & 4825 & 120 & 60 & 141 & 141 \\
\hline\hline
\end{tabular}}
\caption{
Finite-size cumulative depths for the weakly thermalizing Ising state $|X+\rangle$. For each threshold, the columns report the occupation depth, the three active diagnostic depths, and the enclosing memory-core depth. Comparing $q=0.95$ and $q=0.99$ tests the sensitivity of the inferred shallow core to the cumulative tails.
}
\label{tab:app_ising_core_scaling}
\end{table*}

At $q=0.95$, the memory-core depth remains between $24$ and $36$
over the accessible sizes, while the occupation depth grows from $205$ to
$2053$. At $q=0.99$, the memory-core depth varies nonmonotonically as
\[
n_{\rm mc}^{(0.99)}=98,\;97,\;121,\;160,\;141
\]
for $L=10,\ldots,14$, whereas $n_\Pi^{(0.99)}$ grows from $384$ to $4825$.
At the largest size,
\begin{equation}
 n_{\rm mc}^{(0.99)}/\mathcal D_0 \approx 0.017,
 \qquad
 n_\Pi^{(0.99)}/n_{\rm mc}^{(0.99)} \approx 34.2.
\end{equation}
The corresponding tail-induced increases are
\begin{equation}
 \Delta n_{\rm mc}^{(99-95)}=109,
 \qquad
 \Delta n_\Pi^{(99-95)}=2772.
\end{equation}
The shallow core--halo hierarchy therefore persists under the stricter threshold. Over the accessible sizes, $n_{\rm mc}$ shows no systematic growth comparable to that of either $\mathcal D_0$ or $n_\Pi$, providing no finite-size evidence here for an anomalously growing memory core.

\subsection{Confinement-sensitive bubble state}

We next consider the bubble state in the confinement regime,
\begin{equation}
 J=1,
 \qquad
 g=0.5,
 \qquad
 h=0.1.
\end{equation}
The cyclic dimension again lies close to, but need not coincide with, the reflection-even sector dimension.

\begin{table*}[t]
\centering
\tiny
\setlength{\tabcolsep}{4.0pt}
\resizebox{\textwidth}{!}{
\begin{tabular}{c c c | c c c c c | c c c c c}
\hline\hline
&&&
\multicolumn{5}{c|}{$q=0.95$}
&
\multicolumn{5}{c}{$q=0.99$}
\\
$L$
& $\dim\mathcal H^{\rm Ising}$
& $\mathcal D_0$
& $n_\Pi^{(q)}$
& $n_\chi^{(q)}$
& $n_d^{(q)}$
& $n_\Gamma^{(q)}$
& $n_{\rm mc}^{(q)}$
& $n_\Pi^{(q)}$
& $n_\chi^{(q)}$
& $n_d^{(q)}$
& $n_\Gamma^{(q)}$
& $n_{\rm mc}^{(q)}$
\\
\hline
10 & 1024  & 528  & 190 & 71 & 75 & 55 & 75 & 326 & 120 & 295 & 78 & 295 \\
11 & 2048  & 1056 & 328 & 85 & 78 & 59 & 85 & 594 & 133 & 283 & 90 & 283 \\
12 & 4096  & 2079 & 468 & 106 & 83 & 84 & 106 & 960 & 166 & 287 & 126 & 287 \\
13 & 8192  & 4159 & 738 & 123 & 103 & 110 & 123 & 1639 & 213 & 273 & 156 & 273 \\
14 & 16384 & 8244 & 1159 & 144 & 116 & 129 & 144 & 2815 & 302 & 163 & 207 & 302 \\
\hline\hline
\end{tabular}}
\caption{
Finite-size cumulative depths for the confinement-sensitive bubble state. The $99\%$ columns test whether the broader confinement-induced core identified at $q=0.95$ remains separated from both the stationary occupation halo and the cyclic dimension when the cumulative tails are included more stringently.
}
\label{tab:app_confinement_core_scaling}
\end{table*}

At $q=0.95$, $n_{\rm mc}^{(0.95)}$ grows from $75$ at $L=10$ to
$144$ at $L=14$, while $n_\Pi^{(0.95)}$ grows from $190$ to $1159$. At
$q=0.99$, the memory-core depth varies nonmonotonically as
\[
n_{\rm mc}^{(0.99)}=295,\;283,\;287,\;273,\;302,
\]
whereas the occupation depth increases from $326$ to $2815$. At the largest
size,
\begin{equation}
 \Delta n_{\rm mc}^{(99-95)}=158,
 \qquad
 \Delta n_\Pi^{(99-95)}=1656,
\end{equation}
and
\begin{equation}
 n_{\rm mc}^{(0.99)}/\mathcal D_0 \approx 0.037,
 \qquad
 n_\Pi^{(0.99)}/n_{\rm mc}^{(0.99)} \approx 9.32.
\end{equation}
The confinement-induced core remains well separated from the cyclic dimension at both thresholds. It also remains substantially smaller than the stationary occupation background, although this separation is weaker than in the weakly thermalizing Ising state. The monotonic growth of the $95\%$ memory-core depth over the accessible sizes is consistent with the anomalous, confinement-sensitive character of this state.

\subsection{Scarred PXP N\'eel state}

Finally, we consider the periodic PXP chain initialized in $|Z_2\rangle$. The full constrained Hilbert-space dimension is
\begin{equation}
 \dim\mathcal H_{\rm PXP}^{\rm per}(L)
 =
 F_{L-1}+F_{L+1},
\end{equation}
where $F_m$ denotes the Fibonacci sequence. The numerically resolved cyclic dimension $\mathcal D_0$ is obtained from the distinct active energy groups retained at the stated $\varepsilon_E$ and $\varepsilon_\omega$, following the procedure described in Appendix~\ref{app:pxp_dimension_count}. The values used here are those established in Appendix~\ref{app:pxp_dimension_count}:
\begin{equation}
 \mathcal D_0 = 853,\;1937,\;4569,\;10793,\;26015
\end{equation}
for $L=20,22,24,26,28$, respectively. Thus $\mathcal D_0$ is neither the full constrained Hilbert-space dimension nor the dimension of a symmetry sector or forward-scattering truncation.

\begin{table*}[t]
\centering
\tiny
\setlength{\tabcolsep}{4.0pt}
\resizebox{\textwidth}{!}{
\begin{tabular}{c c c | c c c c c | c c c c c}
\hline\hline
&&&
\multicolumn{5}{c|}{$q=0.95$}
&
\multicolumn{5}{c}{$q=0.99$}
\\
$L$
& $\dim\mathcal H_{\rm PXP}^{\rm per}$
& $\mathcal D_0$
& $n_\Pi^{(q)}$
& $n_\chi^{(q)}$
& $n_d^{(q)}$
& $n_\Gamma^{(q)}$
& $n_{\rm mc}^{(q)}$
& $n_\Pi^{(q)}$
& $n_\chi^{(q)}$
& $n_d^{(q)}$
& $n_\Gamma^{(q)}$
& $n_{\rm mc}^{(q)}$
\\
\hline
20 & 15127  & 853   & 141  & 32 & 52  & 21  & 52  & 294   & 76  & 691  & 61  & 691  \\
22 & 39603  & 1937  & 252  & 44 & 46  & 31  & 46  & 883   & 104 & 1306 & 84  & 1306 \\
24 & 103682 & 4569  & 895  & 59 & 55  & 44  & 59  & 2672  & 134 & 1916 & 120 & 1916 \\
26 & 271443 & 10793 & 1530 & 78 & 70  & 61  & 78  & 5135  & 184 & 273  & 159 & 273  \\
28 & 710647 & 26015 & 9818 & 109 & 100 & 92  & 109 & 17723 & 245 & 273  & 214 & 273  \\
\hline\hline
\end{tabular}}
\caption{
Finite-size cumulative depths for the scarred PXP N\'eel state. The $95\%$ and $99\%$ columns compare the dominant active region with its far-tail extension. The large-size profiles are evaluated with the chiral-pair-resolved formulas; $\varepsilon_\Delta$ enters only the finite-size validation at $L=18,20,22,24$ described in Appendix~\ref{app:pxp_gap_validation}. The values of $\mathcal D_0$ are those established in Appendix~\ref{app:pxp_dimension_count}.
}
\label{tab:app_pxp_core_scaling}
\end{table*}

At $q=0.95$, the memory-core depth increases from $52$ at $L=20$ to $109$ at $L=28$, whereas the occupation depth grows to $9818$. At $q=0.99$,
\begin{equation}
 n_{\rm mc}^{(0.99)}
 =
 691,\;1306,\;1916,\;273,\;273
\end{equation}
for $L=20,22,24,26,28$, respectively. This sequence is strongly nonmonotonic: the large values at $L=22$ and $L=24$ are set by the $d_n$ contribution, which changes substantially relative to the other diagnostics between $L=24$ and $L=26$. The $99\%$ data therefore serve as a stringent robustness test of whether the core--halo hierarchy survives when the far tails are included, rather than as evidence for a monotonic finite-size scaling law.

At the largest accessible size $L=28$,
\begin{equation}
 n_{\rm mc}^{(0.99)}/\mathcal D_0 \approx 0.0105,
 \qquad
 n_\Pi^{(0.99)}/n_{\rm mc}^{(0.99)} \approx 64.9,
\end{equation}
while
\begin{equation}
 \Delta n_{\rm mc}^{(99-95)}=164,
 \qquad
 \Delta n_\Pi^{(99-95)}=7905.
\end{equation}
At $L=28$, the occupation depth exceeds the memory-core depth by factors of approximately $90$ and $65$ at $95\%$ and $99\%$, respectively. The strong scarred core--halo separation observed at the primary threshold thus remains present at the largest accessible size when the stricter threshold is applied.

\subsection{Numerical stability and finite-size interpretation}

Variations of $\varepsilon_E$ and $\varepsilon_\omega$ test the numerical construction of the active spectral measure and the associated cyclic dimension. The separate PXP validation in Appendix~\ref{app:pxp_gap_validation} tests the complete equal-gap calculation under variation of $\varepsilon_\Delta$. These procedures probe distinct numerical effects: $\varepsilon_E$ and $\varepsilon_\omega$ can change the retained active spectrum, whereas $\varepsilon_\Delta$ changes only the grouping of nonzero transition frequencies in the finite-size validation branch. At fixed active spectrum, $\Pi_n$, $d_n$, $n_\Pi^{(q)}$, and $n_d^{(q)}$ are independent of $\varepsilon_\Delta$; only $\overline{\chi_n}$, $\Gamma_n$, and their associated cumulative depths can respond to the gap grouping.

The two-threshold comparison probes a different form of robustness. Increasing $q$ from $0.95$ to $0.99$ leaves the underlying profiles and their integrated weights unchanged; it changes only the cumulative-depth threshold used to quantify the spatial extent of their tails. The finite-size core--halo interpretation is strongest when the hierarchy in Eq.~\eqref{eq:app_two_threshold_hierarchy} persists at both thresholds. For the largest sizes studied here,
\begin{align}
 \frac{n_{\rm mc}^{(0.95)}}{\mathcal D_0}
 &=
 \{0.0039,\;0.0175,\;0.0042\},\cr
 \frac{n_{\rm mc}^{(0.99)}}{\mathcal D_0}
 &=
 \{0.0171,\;0.0366,\;0.0105\},
 \label{eq:app_two_threshold_active_fractions}
\end{align}
for the weakly thermalizing Ising, confinement-sensitive bubble, and scarred PXP sequences, respectively. The corresponding halo ratios $n_\Pi^{(q)}/n_{\rm mc}^{(q)}$ are
\begin{equation}
 \{64.2,\;8.05,\;90.1\}
\end{equation}
at $q=0.95$ and
\begin{equation}
 \{34.2,\;9.32,\;64.9\}
\end{equation}
at $q=0.99$.

The present system sizes do not determine a unique asymptotic exponent.
The appropriate finite-size conclusion is therefore restricted to the
observed hierarchy. At the largest accessible size of each sequence, the
memory-related region is much smaller than the cyclic space and remains
embedded within a substantially broader stationary occupation background at
both thresholds. At the primary $95\%$ threshold, the memory-core depth grows
monotonically over the accessible sizes for the bubble and PXP sequences. For
the weakly thermalizing Ising sequence, by contrast,
$n_{\rm mc}^{(0.95)}$ remains between $24$ and $36$ and shows no systematic
growth over $L=10$--$14$. This comparison demonstrates that core--halo
separation and the finite-size behavior of the core depth are distinct
diagnostics. The present sizes do not establish an asymptotic scaling law
for any of the three sequences.

This distinction is important: core--halo separation by itself is not sufficient evidence for anomalous memory. A shallow core can coexist with a large occupation background even in a thermalizing state. The relevant signature is instead the finite-size behavior of the memory-core depth. Claims of asymptotic $O(L)$ behavior therefore require an independent finite-size scaling analysis; threshold robustness, including comparison of the $95\%$ and $99\%$ sequences, provides an additional consistency check rather than a substitute for that analysis.
\FloatBarrier




\bibliography{literature}

\end{document}